\pdfoutput=1
\PassOptionsToPackage{table,xcdraw,svgnames}{xcolor}
\documentclass[acmsmall,screen,nonacm,table]{acmart}%

\setcopyright{none}

\usepackage{booktabs}   %
\usepackage{subcaption} %

\usepackage[utf8]{inputenc}            %
\usepackage[T1]{fontenc}               %
\usepackage{amsmath,mathtools,amsthm}  %
\usepackage{graphicx}
\usepackage{thmtools}      %
\usepackage{mathpartir}                %
\usepackage{stmaryrd}                  %
\usepackage{xspace}                    %

\usepackage{url}
\usepackage{xcolor}             %
\usepackage{array}
\usepackage{tikz}                      %
\usetikzlibrary{trees,calc}
\usepackage[ruled, noend]{algorithm2e}
\usepackage[normalem]{ulem}
\usepackage{listings}
\usepackage{enumerate}                 %
\usepackage{enumitem}
\usepackage{cleveref}                  %
\usepackage{placeins}         %
\usepackage{threeparttable}
\usepackage{fontawesome}               %
\usepackage{pifont}                    %
\usepackage[scaled=0.85]{beramono}
\DeclareMathAlphabet{\mathbbm}{U}{bbm}{m}{n}
\usepackage{hyperref}
\hypersetup{pdfborder={0 0 0}}
\usepackage{pdfpages}
\usepackage{pax}
\usepackage{halloweenmath}
\usepackage{trimclip}
\usepackage{scalerel}[2016/12/29]

\DeclareRobustCommand{\lightning}{%
  \ifmmode
    \text{\fontencoding{U}\fontfamily{wasy}\selectfont\symbol{18}}%
  \else
    {\fontencoding{U}\fontfamily{wasy}\selectfont\symbol{18}}%
  \fi
}

\DeclareMathAlphabet{\mathbsf}{\encodingdefault}{\sfdefault}{bx}{n}

\makeatletter
\newcommand{\showlineheight}{{\color{blue}Current line height: \the\baselineskip}}
\newcommand{\showfontsize}{{\color{blue} [Current font size: \f@size pt]}}
\newcommand{\showall}{{\color{blue} [font size: \f@size pt, line height: \the\baselineskip]}}
\makeatother

\newcommand{\LangName}[1]{\text{#1}}

\newcommand{\Calc}{\Fresco}

\newcommand{\Fresco}{\LangName{Fresco}\xspace}

\newcommand{\FreezeML}{\LangName{FreezeML}\xspace}

\newcommand{\Met}{\textsf{\textsc{Met}}\xspace}

\newcommand*\mynote[3]
{%
  \ifvmode
    \textcolor{#2}{#1: #3}
  \else
    \textcolor{#2}{~#1: #3}
  \fi
}

\definecolor{codegreen}{rgb}{0,0.6,0}
\definecolor{codeblue}{rgb}{0,0,0.8}
\definecolor{codegray}{rgb}{0.4,0.4,0.4}
\definecolor{codepurple}{rgb}{0.58,0,0.82}
\definecolor{codeone}{HTML}{9B26B6}
\definecolor{codetwo}{HTML}{414C87}
\definecolor{backcolour}{rgb}{0.95,0.95,0.92}

\definecolor{myred}{HTML}{BB0000}
\definecolor{myblue}{HTML}{003399}
\definecolor{dred}{HTML}{EB212E}
\definecolor{dblue}{HTML}{2E67F8}
\definecolor{lblue}{HTML}{00B5E2}

\newcommand{\red}[1]{{\color{myred}{#1}}}

\newcommand{\dblue}[1]{{\color{dblue}{#1}}}

\newcommand{\gray}[1]{{\color{codegray}{#1}}}
\newcommand{\black}[1]{{\color{black}{#1}}}

\colorlet{hlcolor}{lightgray}
\newcommand{\highlightwithstyle}[2]{
  {\setlength{\fboxsep}{2pt} \pgfsetfillopacity{0.3} \colorbox{hlcolor}{\pgfsetfillopacity{1}$#1#2$}}
}
\newcommand{\hl}[1]{\mathpalette\highlightwithstyle{#1}}

\makeatletter
\newcommand{\sigb}[2]{
  \@ifmtarg{#1}{:}{:^{#1}} #2
}
\makeatother

\makeatletter
\newcommand{\ogeneric}[2][0.7]{%
  \vphantom{{\oplus}}\mathpalette\o@generic{{#1}{#2}}%
}
\newcommand{\o@generic}[2]{\o@@generic#1#2}
\newcommand{\o@@generic}[3]{%
  \begingroup
  \sbox\z@{$\m@th#1\oplus$}%
  \dimen@=\dimexpr\ht\z@+\dp\z@\relax
  \savebox\tw@[\totalheight]{$\m@th#1\bigcirc$}%
  \makebox[\wd\z@]{%
    \ooalign{%
      $#1\vcenter{\hbox{\resizebox{\dimen@}{!}{\usebox\tw@}}}$\cr
      \hidewidth
      $#1\vcenter{\hbox{\resizebox{#2\dimen@}{!}{$#1\vphantom{\oplus}{#3}$}}}$%
      \hidewidth
      \cr
    }%
  }%
  \endgroup
}
\makeatother

\newcommand{\meta}[1]{\mathsf{#1}}

\definecolor{cred}{HTML}{E10600}
\definecolor{cblue}{HTML}{0077BB}
\definecolor{ccyan}{HTML}{33BBEE}
\definecolor{corange}{HTML}{EE7733}
\definecolor{cteal}{HTML}{009988}
\definecolor{cmagenta}{HTML}{EE3377}
\definecolor{cpurple}{HTML}{AA3377}
\definecolor{cgrey}{HTML}{999999}
\newcommand{\syn}[1]{
  {\color{cred} #1}
}
\newcommand{\inh}[1]{
  {\color{cblue} #1}
}

\newcommand{\gus}[1]{
  \ifx#1\to\mathrel{\color{corange} #1}
  \else{\color{corange} #1}\fi
}
\newcommand{\bid}[1]{
  {\color{cgrey} #1}
}
\newcommand{\inv}[1]{
  \ifx#1\to\mathrel{\color{cpurple} #1}
  \else{\color{cpurple} #1}\fi
}
\newcommand{\noc}[1]{
  \ifx#1\to\mathrel{\color{black} #1}
  \else{\color{black} #1}\fi
}
\newcommand{\col}[1]{
  \ifx#1\to\mathrel{\color{cgrey} #1}
  \else{\color{cgrey} #1}\fi
}
\newcommand{\app}[1]{
  \ifx#1\to\mathrel{\color{corange} #1}
  \else{\color{corange} #1}\fi
}

\newcommand{\sigentails}[1]{\mathbin{[{\text{\footnotesize ${#1}$}}]\!\text{-\!-}}\,}

\newcommand{\tunit}{1}

\newcommand{\eunit}{()}

\let\BoxSym\Box
\newcommand{\squareop}[1]{%
  {\mathrel{\ooalign{\hss\raise-0.1ex\hbox{\scalebox{1.1}{$\BoxSym$}}\hss\cr%
  \kern0.3ex\raise0.12ex\hbox{\scalebox{0.65}{$#1$}}}}}
}

\makeatletter
\DeclareFontEncoding{LS2}{}{\noaccents@}
\DeclareFontSubstitution{LS2}{stix2}{m}{n}
\DeclareSymbolFont{arrows3}{LS2}{stixtt}{m}{n}
\DeclareMathSymbol{\squarelrblackbin}{\mathord}{arrows3}{"89}

\DeclareFontEncoding{LS1}{}{}
\DeclareFontSubstitution{LS1}{stix}{m}{n}
\DeclareSymbolFont{symbols2}{LS1}{stixfrak}{m}{n}
\DeclareMathSymbol{\typecolon}{\mathbin}{symbols2}{"25}
\makeatother

\makeatletter
\newcommand{\Scale}[2][1]{\scalebox{#1}{$\m@th#2$}}
\makeatother

\makeatletter
\newcommand{\varb}[2]{
  \@ifmtarg{#1}{:}{:_{\dblue{#1}}} #2
}
\newcommand{\Letm}[2]{
  \@ifmtarg{#1}{\keyw{let}}{\keyw{let}_{#1}} \; \keyw{mod}_{#2}\;
}
\newcommand{\Letmf}[2]{
  \@ifmtarg{#1}{\keyw{let}}{\keyw{let}_{#1}} \; #2\;
}

\newcommand{\Casem}[1]{
  \@ifmtarg{#1}{\keyw{case}}{\keyw{case}_{#1}} \;
}
\makeatother

\renewcommand{\Box}{\keyw{mod}}

\newcommand{\aex}[1]{{\langle #1\rangle}}

\newcommand{\aeq}[1]{[#1]}

\newcommand{\aconst}[1]{\red{dontuseme}}

\newcommand{\aremove}[1]{\red{dontuseme}}

\makeatletter
\newcommand{\superimpose}[2]{{%
  \ooalign{%
    \hfil$\m@th#1\@firstoftwo#2$\hfil\cr
    \hfil$\m@th#1\@secondoftwo#2$\hfil\cr
  }%
}}
\makeatother

\newcommand{\To}{\Rightarrow}

\makeatletter
\newcommand{\oset}[3][0ex]{%
  \mathrel{\mathop{#3}\limits^{
    \vbox to#1{\kern-2\ex@
    \hbox{$\scriptstyle#2$}\vss}}}}
\makeatother

\newcommand{\guarded}[1]{#1\text{ guarded}}

\newcommand{\nonflex}[1]{#1\text{ non-flex}}

\InputIfFileExists{dotted-ghost.tex}{}{%
  \input{../dotted-ghost.tex}%
}
\newcommand{\DottedHole}{{\mydottedghost}}
\newcommand{\DottedTopHole}{\dottedforallghost}

\makeatletter
\newcommand*\forallghostraise{.16ex} %
\newcommand*\forallscale{0.4}       %

\newcommand{\forallghost}{\mathord{\mathpalette\forallghost@{}}}
\newcommand{\forallghost@}[2]{%
  \ooalign{%
    \hfil$\m@th#1\mathpalette\raiseghost\relax$\hfil\cr
    \hfil\raise\forallghostraise\hbox{%
      \scalebox{\forallscale}{$\m@th#1\forall$}}%
    \hfil\cr
  }%
}

\newcommand*\otherghostraise{.16ex} %

\newcommand{\inferghost}{\mathord{\mathpalette\inferghost@{}}}
\newcommand{\inferghost@}[2]{%
  \ooalign{%
    \hfil$\m@th#1\mathpalette\raiseghost\relax$\hfil\cr
    \hfil\raise\otherghostraise\hbox{%
      \scalebox{\forallscale}{$\m@th#1 \mathtt{I}$}}%
    \hfil\cr
  }%
}

\newcommand{\checkghost}{\mathord{\mathpalette\checkghost@{}}}
\newcommand{\checkghost@}[2]{%
  \ooalign{%
    \hfil$\m@th#1\mathpalette\raiseghost\relax$\hfil\cr
    \hfil\raise\otherghostraise\hbox{%
      \scalebox{\forallscale}{$\m@th#1 \mathtt{C}$}}%
    \hfil\cr
  }%
}

\newcommand{\noguessghost}{\mathord{\mathpalette\noguessghost@{}}}
\newcommand{\noguessghost@}[2]{%
  \ooalign{%
    \hfil$\m@th#1\mathpalette\raiseghost\relax$\hfil\cr
    \hfil\raise\otherghostraise\hbox{%
      \scalebox{\forallscale}{$\m@th#1 \mathtt{N}$}}%
    \hfil\cr
  }%
}
\makeatother

\newcommand{\Hole}{{\myghost}}
\newcommand{\TopHole}{\forallghost} %

\newcommand{\myghost}{\mathpalette\raiseghost\relax}
\newcommand{\raiseghost}[2]{\raisebox{.5\depth}{\scalebox{0.9}{$#1\mathghost$}}}

\newcommand{\foralltag}[1]{\forall^{#1}}

\newcommand{\HoleG}{\DottedHole}
\newcommand{\TopHoleG}{\DottedTopHole}

\definecolor{tagcol}{HTML}{009A17}

\newcommand{\Log}{\code{log}}

\makeatletter
\newcommand*{\@rowstyle}{}

\newcommand*{\rowstyle}[1]{%
  \gdef\@rowstyle{#1}%
  \@rowstyle\ignorespaces%
}

\newcolumntype{=}{%
  >{\gdef\@rowstyle{}}%
}

\newcolumntype{+}{%
  >{\@rowstyle}%
}
\makeatother

\newcommand{\Freeze}[1]{\lceil #1 \rceil}

\newcommand{\M}{\mathtt{m}}
\newcommand{\N}{\mathtt{n}}
\newcommand{\TY}{\mathtt{ty}}

\newcommand{\PI}{\mathtt{sk}}
\newcommand{\SK}{\mathtt{sk}}

\DeclareDocumentCommand{\FN}{ o }{
\IfNoValueTF{#1}{\mathtt{F}\,}{\mathtt{F}^{#1}\,}
}

\newcommand{\TI}{\mathtt{I}}
\newcommand{\TC}{\mathtt{C}}

\newcommand{\gmargin}[1]{\mkern6mu #1 \mkern6mu}

\newcommand{\gor}{\gmargin{\text{\textcolor{gray}{or}}}}

\newcommand{\TTo}{\mathrel{\mathrlap{\Shortto}\phantom{~}\!\Shortto}}

\makeatletter
\DeclareRobustCommand{\Shorteq}{%
  \mathrel{\mathpalette\Short@eq\relax}%
}
\newcommand{\Short@eq}[2]{%
  \mkern2mu
  \clipbox{{.4\width} 0 0 0}{$\m@th#1\vphantom{+}{\Relbar}$}%
  }

\DeclareRobustCommand{\Shortto}{%
  \mathrel{\mathpalette\Short@to\relax}%
}
\newcommand{\Short@to}[2]{%
  \mkern2mu
  \clipbox{{.4\width} 0 0 0}{$\m@th#1\vphantom{+}{\Rightarrow}$}%
  }

\DeclareRobustCommand{\Shortfrom}{%
  \mathrel{\mathpalette\Short@from\relax}%
}
\newcommand{\Short@from}[2]{%
  \clipbox{0 0 {.4\width} 0}{$\m@th#1\vphantom{+}{\Leftarrow}$}%
  \mkern2mu
  }

\DeclareRobustCommand{\Shortfto}{%
  \mathrel{\mathpalette\Short@fto\relax}%
}
\newcommand{\Short@fto}[2]{%
  \mathrel{\mathrlap{\Shortfrom}\mkern-4mu\Shortto}%
  }
\makeatother

\newcommand{\jdghl}[1]{#1}
\newcommand{\skhl}[1]{{\color{cblue}{#1}}}

\DeclareDocumentCommand{\jdg}{ o o o m m }{
  #4
  \mathrel{\jdghl{
    \IfNoValueTF{#3}{\vdash}{\sigentails{#3}}
  }^{
    \IfNoValueTF{#1}{}{#1}
    }_{
    \IfNoValueTF{#2}{}{#2}
  }}
  #5
}
\DeclareDocumentCommand{\mov}{ o o o m m m }{
  #4
  \mathrel{\jdghl{
    \IfNoValueTF{#3}{\vdash}{\sigentails{#3}}
  }^{
    \IfNoValueTF{#1}{}{#1}
    }_{
    \IfNoValueTF{#2}{}{#2}
  }}
  #5
  \mathrel{\jdghl{\dashv}} #6
}
\DeclareDocumentCommand{\atypp}{ o o m m m m }{
    #3
    \mathrel{\jdghl{\sigentails{#2}}_{
      #1 }}
    #4
    \mathrel{\jdghl{\To}}
    #5
    \mathrel{\jdghl{\dashv}}
    #6
}
\DeclareDocumentCommand{\apatp}{ o o m m m m }{
    #3
    \mathrel{\jdghl{\sigentails{#2}}_{
       #1 }}
    #4
    \mathrel{\jdghl{\TTo}}
    #5
    \mathrel{\jdghl{\dashv}}
    #6
}
\DeclareDocumentCommand{\atypsub}{ o o o m m m m }{%
    #4
    \mathrel{\jdghl{\sigentails{#3}}^{
      #1
    }_{
      #2
    }} #5 \mathrel{\jdghl{\subtype}} #6%
    \mathrel{\jdghl{\dashv}}
    #7
}
\DeclareDocumentCommand{\asubtag}{ o o o o m m m m }{%
    #4\mid
    #5
    \mathrel{\jdghl{\sigentails{#3}^{#1}_{#2}}}
    #6 \mathrel{\jdghl{\subtypedec}} #7%
    \mathrel{\jdghl{\dashv}}
    #8
}
\DeclareDocumentCommand{\aconsis}{ o o m m m m }{%
    #3
    \mathrel{\jdghl{
      \IfNoValueTF{#2}{
        \IfNoValueTF{#1}{
          \vdash%
        }{%
          \sigentails{#1}%
        }%
      }{%
        \sigentails{#2}%
      }_{\IfNoValueTF{#2}{}{#1}}
    }} #4 \mathrel{\jdghl{\consis}} #5
    \mathrel{\jdghl{\dashv}}
    #6
}
\makeatletter
\DeclareDocumentCommand{\aconsistag}{ o o o m m m m }{%
    \IfNoValueTF{#3}{
      \IfNoValueTF{#2}{
        \IfNoValueTF{#1}{}{#1\mid}%
        #4
        \mathrel{\jdghl{\vdash}}%
      }{%
        #2\mid
        #4
        \mathrel{\jdghl{\sigentails{#1}}}%
      }%
    }{
      #3\mid #4
      \mathrel{\jdghl{\sigentails{#2}_{#1}}}%
    }
    #5 \mathrel{\jdghl{\consisdec}} #6
    \mathrel{\jdghl{\dashv}}
    #7
}
\makeatother
\DeclareDocumentCommand{\abroom}{ o o o m m m m }{%
    #4
    \mathrel{\jdghl{\vdash}^{
      #1
    }_{
      #2
    }} #5 \mathrel{\jdghl{\xrightbroom{\black{#3}}}} #6%
    \mathrel{\jdghl{\dashv}}
    #7
}

\DeclareDocumentCommand{\ajoin}{ m m m o m }{%
    #1
    \mathrel{\jdghl{\vdash}}
    \IfNoValueTF{#4}{
    #2 \mathrel{\jdghl{\consis}} #3
      }{
    #2 \mathrel{\jdghl{\consis}} #3
    \mathrel{\jdghl{\To}} #4
    }
    \mathrel{\jdghl{\dashv}}
    #5
}
\newcommand{\typi}[3]{\typp[\TI]{#1}{#2}{#3}}
\newcommand{\typc}[3]{\typp[\TC]{#1}{#2}{#3}}
\newcommand{\pati}[3]{\patp[\TC]{#1}{#2}{#3}}
\newcommand{\ngti}[3]{\patp[\TI]{#1}{#2}{#3}}

\DeclareDocumentCommand{\typp}{ o o m m m }{%
    #3
    \mathrel{%
      \IfNoValueTF{#1}{
        \jdghl{\vdash}
      }{
        \IfNoValueTF{#2}{
          \jdghl{\vdash}_{#1}
        }{
          \jdghl{\vdash}^{#1}_{#2}
        }
      }%
    } #4 \mathrel{\jdghl{:}} #5%
}
\DeclareDocumentCommand{\patp}{ o o o m m m }{%
    #4
    \IfNoValueTF{#3}{
      \IfNoValueTF{#2}{}{\mid \skhl{#2}}
    }{\mid \skhl{#3}}
    \mathrel{\jdghl{\vdash}^{
      \IfNoValueTF{#3}{}{#1}
    }_{
      \IfNoValueTF{#3}{
        #1
      }{#2}
    }} #5 \mathrel{\jdghl{\TTo}} #6%
}
\DeclareDocumentCommand{\typsub}{ o o o m m m }{%
    \IfNoValueTF{#3}{
      \IfNoValueTF{#2}{}{#2\mid}
    }{\if\relax\detokenize{#3}\relax\else#3\mid\fi}
    #4
    \mathrel{\jdghl{\vdash}^{
      \IfNoValueTF{#3}{}{#1}
    }_{
      \IfNoValueTF{#3}{#1}{#2}
    }} #5 \mathrel{\jdghl{\subtype}} #6%
}
\DeclareDocumentCommand{\patsub}{ o o m m m }{%
    \IfNoValueTF{#2}{}{#2\mid}
    #3
    \mathrel{\jdghl{\vdash}^{
      \PI
    }_{
      #1
    }} #4 \mathrel{\jdghl{\subtype}} #5%
}
\DeclareDocumentCommand{\allsub}{ o o m m m }{%
    #3
    \mathrel{\jdghl{\vdash}^{
      #1
    }_{
      #2
    }} #4 \mathrel{\jdghl{\subtype}} #5%
}
\DeclareDocumentCommand{\typconsis}{o m m m}{
  \IfNoValueTF{#1}{}{
    #1 \mid
  }
  #2 \mathrel{\jdghl{\vdash}} #3 \sim #4
}

\newcommand{\stypi}[3]{#1 \mathrel{\jdghl{\vdash}_\TI} #2 : #3}
\newcommand{\stypc}[3]{#1 \mathrel{\jdghl{\vdash}_\TC} #2 : #3}

\DeclareDocumentCommand{\skmod}{o m m}{
  #2 :^{\IfNoValueTF{#1}{}{#1}} #3
}

\DeclareDocumentCommand{\sdtypp}{ o o o m m m }{%
    #4
    \IfNoValueTF{#3}{
      \IfNoValueTF{#2}{}{\mid \skhl{#2}}
    }{\mid \skhl{#3}}
    \mathrel{\jdghl{\vdash}^{\mathrm{s}\IfNoValueTF{#3}{}{,#1}}_{
      \IfNoValueTF{#3}{
        #1
      }{#2}
    }} #5 \mathrel{\jdghl{:}} #6%
}

\DeclareDocumentCommand{\sksem}{o m}{
  \llbracket #2 \rrbracket_{\IfNoValueTF{#1}{}{#1}}
}

\newcommand{\tags}[2]{#2 \mathrel{\#} #1}

\renewcommand{\fill}{\mathrel{\btright}}
\newcommand{\fillsyn}{\mathrel{\syn{\btright}}}
\DeclareDocumentCommand{\skctx}{o m}{
  \IfNoValueTF{#2}{#1 \vdash}{
    #2 \mid #1 \vdash
  }
}
\DeclareDocumentCommand{\split}{ o o m } {
  \IfNoValueTF {#1}
  {\mathrel{\jdghl{\tright}_{#3}}}
  {
  \IfNoValueTF {#2}
  {\mathrel{\jdghl{\tright}_{#3}}}
  {{#1} \mathrel{\jdghl{\tright}_{#3}} {#2}}
  }
}

\DeclareDocumentCommand{\unify}{m m o}{
  #1 \mathrel{\jdghl{\sim}} #2
  \IfNoValueTF{#3}{}{
    \mathrel{\jdghl{\To}} #3
  }
}
\DeclareDocumentCommand{\prejoin}{m m o}{
  \meta{prejoin}(#1; #2)
  \IfNoValueTF{#3}{}{
    \mathrel{\jdghl{\To}} #3
  }
}
\DeclareDocumentCommand{\presub}{m m o}{
  \meta{presub}(#1; #2)
  \IfNoValueTF{#3}{}{
    \mathrel{\jdghl{\To}} #3
  }
}
\newcommand{\tyequiv}[2]{#1 \mathrel{{\equiv}} #2}

\newcommand{\unifyvar}[2]{#1 \mathrel{\jdghl{\simeq}} #2}
\newcommand{\assignvar}[2]{#1 \mathrel{\jdghl{\coloneq}} #2}

\makeatletter
\newcommand{\solvetag}[4]{
  #3
  \@ifmtarg{#1}{\vdash}{\vdash_{#1}}
  #4 \dashv #2
}
\makeatother

\newcommand{\consis}{\mathrel{\sim}}

\newcommand{\tright}{\mathrel{\raisebox{.5pt}{\scaleobj{0.8}{\vartriangleright}}}}
\DeclareRobustCommand{\btright}{%
  \mathrel{\raisebox{.6pt}{$\scriptstyle\blacktriangleright$}}%
}

\DeclareDocumentCommand{\inc}{o m m}{
  \IfNoValueTF{#1}{}{#1 \mathrel{\jdghl{\dblcolon}}}
  #2 \mathrel{\jdghl{\sqsubseteq}} #3
}

\newcommand{\assign}{\coloneq}
\DeclareDocumentCommand{\decl}{ m o m }{
  #1
  \IfNoValueTF{#2}{:}{\mathrel{\text{\footnotesize $[#2]$}\mkern-6mu:}}
  #3
}

\DeclareDocumentCommand{\defn}{ m o m }{
  #1
  \IfNoValueTF{#2}{\assign}{\mathrel{\text{\footnotesize $[#2]$}\mkern-6mu\assign}}
  #3
}

\makeatletter
\DeclareRobustCommand{\Circle}{%
  \mathbin{\mathpalette\on@ntimes\relax}%
}
\newcommand{\on@ntimes}[2]{%
  \vcenter{\hbox{%
    \sbox0{\m@th$#1\otimes$}%
    \setlength\unitlength{\wd0}%
    \begin{picture}(1,1)
    \linethickness{0.35pt}
    \put(.5,.5){\circle{.8}}
    \end{picture}%
  }}%
}
\makeatother

\newcommand{\txmark}{\text{\textcolor{black}{\ding{51}}{\textcolor{black}{\kern-0.7em\ding{55}}}}}

\newcommand{\subtype}{\mathrel{\leqslant}}
\newcommand{\subtypedec}{\mathrel{\dot{\leqslant}}}
\newcommand{\consisdec}{\mathrel{\dot{\sim}}}

\DeclareDocumentCommand{\precise}{o m m}{
  \IfNoValueTF{#1}{#2 \preceq #3}{#2 \preceq_{#1} #3}
}

\DeclareDocumentCommand{\looksub}{o o m m m o m o}{
  #4 \subtype^{\IfNoValueTF{#2}{}{#1}}_{
    \IfNoValueTF{#2}{#1}{#2}
  } #5
  \IfNoValueTF{#6}{}{;#6}
  \mathrel{\lightning^{#3}} #8 ; #7
}
\DeclareDocumentCommand{\lookeq}{o m m m o m o}{
  #3 \consis #4
  \IfNoValueTF{#5}{}{;#5}
  \mathrel{\lightning^{#2}} #7 ; #6
}

\DeclareDocumentCommand{\poly}{o m}{
  \meta{poly}_{\IfNoValueTF{#1}{}{#1}}(#2)
}
\DeclareDocumentCommand{\nopoly}{o m}{
  \meta{nopoly}_{\IfNoValueTF{#1}{}{#1}}(#2)
}
\newcommand{\mC}{\mathcal{C}}
\DeclareDocumentCommand{\solve}{m m m}{
  #1 ; #2 \rightsquigarrow #3
}
\DeclareDocumentCommand{\solvesub}{m m m m}{
  #1 \subtype_{#2} #3 \rightsquigarrow #4
}
\DeclareDocumentCommand{\solveeq}{m m m}{
  #1 \consis #2 \rightsquigarrow #3
}
\newcommand{\fresh}[1]{#1\text{ fresh}}

\DeclareDocumentCommand{\subcons}{m m}{
  \text{sub}_{#1}(#2)
}
\DeclareDocumentCommand{\eqcons}{m}{
  \text{eq}(#1)
}

\DeclareDocumentCommand{\guess}{ o m }{
  \meta{guess}^{\IfNoValueTF{#1}{}{#1}}(#2)
}
\newcommand{\guessMono}[1]{\guess[\TY]{#1}}

\newcommand{\ftv}[1]{\meta{ftv}(#1)}

\DeclareDocumentCommand{\ol}{o m}{
  \overline{#2}^{\IfNoValueTF{#1}{}{#1}}
}

\newcommand{\Fork}{\keyw{fork}}
\NewDocumentCommand{\ForkC}{ m m O{dual} }{\Fork^{#2}_{#1}} %

\newcommand{\erase}[1]{\lfloor #1 \rfloor}

\DeclareDocumentCommand{\solvetag}{m m m}{
  #1 ; #2 \rightsquigarrow #3
}

\newcommand{\hlc}[2]{%
  \begingroup
  \setlength{\fboxsep}{1pt}%
  \colorbox{#1!10}{\ensuremath{#2}}%
  \endgroup\xspace
}

\newcommand{\circnum}[1]{%
  \begingroup
  \setlength{\unitlength}{1ex}%
  \raisebox{-0.55ex}{%
    \begin{picture}(2.2,2.2)
      \put(1.1,1.1){\circle{2.1}}
      \put(1.1,1.1){\makebox(0,0){\scriptsize #1}}
    \end{picture}%
  }%
  \endgroup
}

\DeclareDocumentCommand{\refo}{o}{{\IfNoValueTF{#1}{}{\ensuremath{#1}}}}
\DeclareDocumentCommand{\refn}{o o}{%
  \hlc{gray}{\IfNoValueTF{#2}{}{#2\;} \IfNoValueTF{#1}{}{\circnum{#1}}}%
}
\DeclareDocumentCommand{\refa}{o}{\hlc{red}{\IfNoValueTF{#1}{ \circnum{1} }{#1\;\circnum{1}}}}
\DeclareDocumentCommand{\refb}{o}{\hlc{blue}{\IfNoValueTF{#1}{ \circnum{2} }{#1\;\circnum{2}}}}
\DeclareDocumentCommand{\refc}{o}{\hlc{orange}{\IfNoValueTF{#1}{ \circnum{3} }{#1\;\circnum{3}}}}
\DeclareDocumentCommand{\refd}{o}{\hlc{purple}{\IfNoValueTF{#1}{ \circnum{4} }{#1\;\circnum{4}}}}
\DeclareDocumentCommand{\refe}{o}{\hlc{green}{\IfNoValueTF{#1}{ \circnum{5} }{#1\;\circnum{5}}}}
\DeclareDocumentCommand{\reff}{o}{\hlc{cyan}{\IfNoValueTF{#1}{ \circnum{6} }{#1\;\circnum{6}}}}
\DeclareDocumentCommand{\refg}{o}{\hlc{brown}{\IfNoValueTF{#1}{ \circnum{7} }{#1\;\circnum{7}}}}
\DeclareDocumentCommand{\refh}{o}{\hlc{yellow}{\IfNoValueTF{#1}{ \circnum{8} }{#1\;\circnum{8}}}}
\DeclareDocumentCommand{\refi}{o}{\hlc{magenta}{\IfNoValueTF{#1}{ \circnum{9} }{#1\;\circnum{9}}}}

\renewcommand{\descriptionlabel}[1]{\hspace{\labelsep}\textrm{\color{gray!85!black}#1}}

\newcommand{\notsmall}{}

\newcommand{\slab}[1]{\textrm{#1}}

\newcommand{\dtylab}[1]{\text{\scshape{D-#1}}}
\newcommand{\sdtylab}[1]{\text{\scshape{SD-#1}}}
\newcommand{\tilab}[1]{\text{\scshape{I-#1}}}
\newcommand{\eqlab}[1]{\text{\scshape{C-#1}}}

\newcommand{\pilab}[1]{\text{\scshape{PI-#1}}}
\newcommand{\sublab}[1]{\text{\scshape{S-#1}}}
\newcommand{\silab}[1]{\text{\scshape{SI-#1}}}
\newcommand{\lslab}[1]{\text{\scshape{LS-#1}}}
\newcommand{\lelab}[1]{\text{\scshape{LE-#1}}}

\newcommand{\solsublab}[1]{\text{\scshape{SolS-#1}}}
\newcommand{\soleqlab}[1]{\text{\scshape{SolE-#1}}}

\newcommand{\plab}[1]{\text{\scshape{P-#1}}}
\newcommand{\ulab}[1]{\text{\scshape{U-#1}}}

\newcommand{\dec}[1]{\code{#1}}

\newcommand{\keyw}[1]{{{\mathbsf{#1}}}}

\renewcommand{\Case}{\keyw{case}\;}
\newcommand{\Casey}{\keyw{case}}
\newcommand{\Of}{\;\keyw{of}\;}

\newcommand{\Pair}[2]{#1 \times #2}

\makeatletter
\DeclareRobustCommand{\circbullet}{\mathbin{\vphantom{\circ}\text{\circbullet@}}}
\newcommand{\circbullet@}{%
  \check@mathfonts
  \m@th\ooalign{%
    \clipbox{0 0 0 {\dimexpr\height-\fontdimen22\textfont2}}{$\bullet$}\cr
    $\circ$\cr
  }%
}
\DeclareRobustCommand{\bulletcirc}{\mathbin{\text{\bulletcirc@}}}
\newcommand{\bulletcirc@}{%
  \check@mathfonts
  \m@th\ooalign{%
    \raisebox{\fontdimen22\textfont2}{\clipbox{0 {\fontdimen22\textfont2} 0 0}{$\bullet$}}\cr
    $\circ$\cr
  }%
}
\makeatother

\newcommand{\code}[1]{{\scalebox{0.94}{\texttt{#1}}}}

\newcommand{\TUnit}{1}
\newcommand{\Int}{\code{Int}}
\newcommand{\Bool}{\code{Bool}}

\newcommand{\List}{\code{List}}

\newcommand{\True}{\mathsf{true}}

\newcommand{\sto}{\twoheadrightarrow}

\newcommand{\ba}{\begin{array}}
\newcommand{\ea}{\end{array}}

\newcommand{\bl}{\ba[t]{@{}l@{}}}
\newcommand{\el}{\ea}

\renewenvironment{displaymath}{\notsmall\[}{\]\normalsize\ignorespacesafterend}

\newenvironment{mathparshrink}
{\begin{@empty}%
    \begin{mathpar}}
      {\end{mathpar}\end{@empty}\ignorespacesafterend}

\newcommand{\codesize}{\fontsize{8.7}{10.4}}
\newcommand{\rulesize}{\fontsize{9}{11.16}}

\lstdefinestyle{ran}{
    commentstyle=\color{codegray},
    numberstyle=\tiny,
    stringstyle=\ttfamily,
    basicstyle=\codesize\selectfont\ttfamily,
    breakatwhitespace=false,
    breaklines=true,
    captionpos=b,
    keepspaces=true,
    numbers=none,
    numbersep=5pt,
    showspaces=false,
    showstringspaces=false,
    showtabs=false,
    tabsize=2,
    columns=fullflexible,
    aboveskip=.9\medskipamount,
    belowskip=.9\medskipamount,
    keywords=[1]{
      do, mask, handle, with, let, in, resume, return, eff, type,
      fst, snd, if, then, else, case, of, fun, raise, maska, data
    },
    keywordstyle=[1]\bfseries,
    keywords=[2]{
      Int, List, Maybe, String, Bool, Pure, Any, Proc, Abs,
      true, false, nil, cons, just, nothing, proc
    },
    keywordstyle=[2]\color{codegreen},
    keywords=[3]{
      fail, get, put, ufork, fork, yield, suspend, abort, log, throw, ask, foo, leak,
      bar, baz, Yield, Gen, State, Fork, Coop, Queue, UCoop
    },
    keywordstyle=[3]\color{dblue},
    morecomment = [l]{\#},
    literate={
      {->}{{$\to$}}{2}
      {=>}{{$\Rightarrow$}}{2}
      {>>}{{$\sto$}}{2}
      {~>}{{$\mapsto$}}{2}
      {\#>}{{$\sharparrow$}}{2}
      {<=}{{$\Leftarrow$}}{1}
      {|-}{{$\vdash$}}{1}
      {|/-}{{$\nvdash$}}{1}
      {@}{{$\gray{\texttt{@}}$}}{1}
      {forall}{{$\forall$}}{1}
      {Unit}{{{\color{codegreen}1}}}{1}
      {Gamma}{{{$\Gamma$}}}{1}
      {earr}{{$\xrightarrow{\texttt{e}}$}}{2}
      {earrg}{{$\xrightarrow{\texttt{\leff{Gen} \ltype{Int}, e}}$}}{2}
    },
    escapeinside={<@}{@>}
}

\newcommand{\sharparrow}{\mathrel{\mkern3mu\raisebox{-.1ex}{\scalebox{1}[1]{\#}}\mkern-17mu\To}}

\newcommand{\leff}[1]{{\color{dblue}#1}}
\newcommand{\ltype}[1]{{\color{codegreen}#1}}

\lstdefinestyle{koka}{
    commentstyle=\color{codegray},
    numberstyle=\tiny,
    stringstyle=\ttfamily\small,
    basicstyle=\codesize\selectfont\ttfamily,
    breakatwhitespace=false,
    breaklines=true,
    captionpos=b,
    keepspaces=true,
    numbers=none,
    numbersep=5pt,
    showspaces=false,
    showstringspaces=false,
    showtabs=false,
    tabsize=2,
    columns=fullflexible,
    aboveskip=.9\medskipamount,
    belowskip=.9\medskipamount,
    keywords=[1]{
      do, mask, handle, with, let, in, resume, return, effect,
      fst, snd, if, then, else, case, of,
    },
    keywordstyle=[1]\bfseries,
    keywords=[2]{
      List, Int, Unit, int, Tuple2
    },
    keywordstyle=[2]\color{codegreen},
    keywords=[3]{
      choose, fail, get, put, ufork, fork, yield, suspend, abort, log, throw, ask, leak,
      bar, baz, Yield, Gen, State, Fork, Coop, gen
    },
    keywordstyle=[3]\color{dblue},
    morecomment = [l]{\#},
    literate={
      {=>}{{$\Rightarrow$}}{2}
      {->}{{$\to$}}{2}
    }
}

\input{appendix-labels}

\begin{document}

\title{Bidirectional Typing with Freezing, Skeletons, and Ghosts}
\subtitle{A New Approach to Sound and Complete Type Inference for First-Class Polymorphism}

\author{Wenhao Tang}
\orcid{0009-0000-6589-3821}
\email{wenhao.tang@ed.ac.uk}
\affiliation{%
  \institution{The University of Edinburgh}
  \country{UK}
}

\author{Shengyi Jiang}
\orcid{0000-0002-4443-0753}
\email{syjiang@cs.hku.hk}
\affiliation{%
  \institution{The University of Hong Kong}
  \country{China}
}

\author{Aghilas Y. Boussaa}
\orcid{0009-0005-5954-1711}
\email{aghilas.boussaa@ens.fr}
\affiliation{
  \institution{École normale supérieure -- PSL}
  \country{France}
}

\author{Sam Lindley}
\orcid{0000-0002-1360-4714}
\email{sam.lindley@ed.ac.uk}
\affiliation{%
  \institution{The University of Edinburgh}
  \country{UK}}

\author{Bruno C. d. S. Oliveira}
\orcid{0000-0002-1846-7210}
\email{bruno@cs.hku.hk}
\affiliation{%
  \institution{The University of Hong Kong}
  \country{China}
}

\begin{abstract}
  Bidirectional typing makes use of local information flow between
  functions and arguments.
  Conventional bidirectional typing only supports unidirectional
  information flow, typically from functions to arguments, which is
  insufficient to infer first-class polymorphism.
  Existing work on improving information flow either has limited
  support for mixed information flow or requires ad hoc mechanisms
  that harm predictability.

  We propose \Calc, a novel bidirectional type inference approach to
  first-class polymorphism.
  \Calc enables local type information to flow back and forth
  between functions and arguments via \emph{skeletons}, with
  \emph{ghosts} representing unknown type information, and allows
  users to customise the direction of information flow by
  \emph{freezing}.
  The flexible information flow of \Calc enables expressive and
  predictable inference for first-class polymorphism.
  We provide a declarative specification for \Calc, a simple type
  inference algorithm that is sound and complete with respect to the
  declarative system, and a prototype implementation that further
  generalises \Calc to infer modal effect types.
\end{abstract}

\maketitle

\section{Introduction}
\label{sec:introduction}

\newcommand{\ChurchList}{\dec{ChurchList}}
\newcommand{\churchMap}{\dec{churchMap}}
\newcommand{\churchIds}{\dec{churchIds}}

Consider the following application of a map function to a Church-encoded list:
\[
\churchMap\;(\lambda f . (f\; 42, f\; \True))\;\churchIds
\]
where $\churchMap$ and $\churchIds$ have the standard types
\[\bl
\churchMap : \forall a\, b . (a \to b) \to \ChurchList\;a \to \ChurchList\;b \\
\churchIds : \ChurchList\;(\forall a . a \to a)
\el\]
with the type alias $\ChurchList\;a = \forall b. b \to (a\to
b\to b) \to b$ for the Church encoding.

This simple example turns out to be surprisingly challenging for many
type inference systems.
We identify two key challenges for type checking our example:
\begin{enumerate}[noitemsep, leftmargin=*]
  \item instantiating the type variable $a$ of $\churchMap$ with the
    polymorphic type $\forall a . a \to a$, and
  \item assigning the polymorphic type $\forall a . a \to a$ to the parameter
    $f$ of the term $\lambda f . (f\; 42, f\; \True)$.
\end{enumerate}
ML-style type inference~\citep{DamasMilner,SML} fails immediately
since it only supports rank-1 polymorphism, which restricts universal
quantifiers to the top level of types, while the types of $\churchMap$
and $\churchIds$ have universal quantifiers nested inside function
arrows.
Bidirectional type inference~\citep{JonesVWS07,DunfieldK13} nicely
supports \emph{higher-rank polymorphism}, allowing quantifiers to
appear flexibly in types, as in the types of $\churchMap$ and
$\churchIds$.
Moreover, bidirectional type inference allows local type information
to flow from functions to arguments, which would address challenge (2)
provided challenge (1) had already been addressed.
However, most existing work on bidirectional type inference with
higher-rank polymorphism is restricted to monomorphic instantiation,
i.e., instantiation with monomorphic types (monotypes), and so does
not address challenge (1), which requires polymorphic instantiation,
i.e., instantiation with polymorphic types (polytypes%
).
We refer to the combination of higher-rank polymorphism \emph{and}
polymorphic instantiation as \emph{first-class polymorphism} (FCP).

Type inference for FCP is a hot topic, enjoying
many different approaches,
including PolyML~\citep{GarrigueR99}, LTI~\citep{PierceT00},
MLF~\citep{BotlanR03}, HML~\citep{Leijen09},
Boxy~\citep{VytiniotisWJ06}, HMF~\citep{Leijen08},
QML~\citep{RussoV09}, GI~\citep{SerranoHVJ18},
QuickLook~\citep{SerranoHJV20}, FCIF~\citep{Kovacs20},
FreezeML~\citep{EmrichLSCC20,EmrichSCL22},
SuperF~\citep{ParreauxBFC24}, LCTI~\citep{XueCJO26}, and
ATIA~\citep{Morris26}.
However, almost all of these approaches still fail to type check the
above example.%
\footnote{The exception is SuperF~\citep{ParreauxBFC24}, which
extends surface type syntax with constraints, but lacks completeness
results.}
Many of these approaches handle polymorphic instantiation
as in challenge (1) by using their own heuristics or constraint solving
algorithms to determine that $\churchMap$ should be instantiated with
the polytype $\forall a . a \to a$ as required by the argument
$\churchIds$.
However, unlike bidirectional type inference, they cannot make use of
local type information flow to address challenge (2).
Some approaches, such as QuickLook~\citep{SerranoHJV20}, the algorithm
behind the \code{ImpredicativeTypes} extension in GHC, do adopt
bidirectional typing techniques to support information flow from
functions to arguments.
This enables QuickLook to address challenge (2), but it fails to
address challenge (1), because it can make only limited use of
information about arguments to guide polymorphic instantiation.

Our two key challenges exemplify the need for different directions of
type information flow between functions and arguments.
To address challenge (1), type information must flow from the second
argument $\churchIds$ to the function $\churchMap$ for polymorphic
instantiation.
To address challenge (2), type information must flow from the
instantiated function to the first argument $\lambda f . (f\; 42, f\;
\True)$ in order to assign a polymorphic type to $f$.
However, combining these two directions of type information flow to
obtain both of their advantages is much more subtle than it may seem.
Several existing type inference systems emphasise mixed directions of
type information
flow~\citep{VytiniotisWJ06,SerranoHJV20,OderskyZZ01,XueCJO26}.
However, as far as we are aware, none of them suffices to type check
the above example due to limitations on how different directions of
type information flow can be mixed.
Moreover, they often rely on ad hoc algorithmic procedures to
propagate local type information from arguments to functions, and
thus lack declarative specifications.

We propose \Calc%
\footnote{%
  The name \Calc is a loose acronym of the paper title ``FREezing,
  Skeletons, and GhOsts''.
  },
a novel type inference approach for FCP that
utilises mixed directions of type information flow.
Our central idea is to allow local type information to first
flow from arguments to functions, addressing challenge (1), and then
flow back from functions to arguments, addressing challenge (2).
Instead of relying on ad hoc mechanisms for collecting local type
information from arguments, \Calc uses \emph{skeleton inference},
which mimics type inference, to infer \emph{skeletons} of arguments.
Skeletons extend types with \emph{ghosts} that represent unknown
information.
An ordinary ghost $\Hole$ represents an unknown type, while a
universal ghost $\TopHole$ represents unknown universal quantifiers.
For instance, a skeleton
$\TopHole.\Hole\to\Int$ %
represents a possibly polymorphic function type with result type $\Int$.
Skeletons enable mixed information flow between functions and
arguments: known information in an argument skeleton is passed to its
function; unknown information denoted by ghosts in the skeleton is
then filled in by the function and passed back to the argument.

In order to provide fine-grained control over which information is
considered known and thus flows from arguments to functions, we
introduce a term-level {freezing} operator inspired by
FreezeML~\citep{EmrichLSCC20}.
Freezing a term fixes its skeleton to be exactly what is inferred from
the term itself.

A fundamental guiding principle of \Calc is
that ``polymorphism should never be guessed out of thin air''.
When instantiating a polymorphic type, \Calc only guesses a universal
quantifier if this quantifier is determined by known information
collected from the environment, such as via skeleton inference.
Inspired by classic ideas on provenance in database
systems~\citep{BunemanKT01} and \citet{Barendregt85}'s proof for
finiteness of developments, we propose a novel approach to formalising
this principle in a declarative specification by tagging each guessed
quantifier.

Unlike many existing approaches to type inference for FCP, \Calc
enjoys a declarative specification.
We prove our type inference algorithm is sound and complete with
respect to the declarative specification.
Moreover, \Calc stays with plain System F types, requiring no
extension of the surface type syntax with constraints. Its
algorithm is simple: unification is used only for monotypes, and
polymorphic instantiation is always resolved locally.
In order to succinctly distinguish synthesised (known from the term)
and inherited (known from the environment) information within a type,
we follow \citet{OderskyZZ01} in ascribing \emph{colours} to types.
This enables us to concisely express the declarative type system in a
form similar to more conventional bidirectional type systems, making
\Calc easier to understand and reason about.

We make no claim that \Calc is the most expressive type inference
system for FCP.
There are examples that type check in other systems but not in \Calc
(and vice versa).
We compare \Calc with several other systems through a range of
examples in \Cref{app:benchmark}.
We believe that \Calc achieves a good balance between
expressiveness and simplicity in the sense that programs that are
obviously well-typed through mixed directions of local type
information between functions and arguments can be type checked in
\Calc.

The main contributions of this paper are as follows.
\begin{itemize}[nosep, leftmargin=*]
  \item We give a high-level overview of \Calc (\Cref{sec:overview}),
  focusing on how \Calc type checks the motivating example from the
  beginning of the paper.
  We demonstrate the key novel ideas of \Calc, including how \Calc
  enables mixed directions of type information flow via skeletons and
  ghosts, and how \Calc avoids guessing polymorphism out of thin air.
  \item We give a declarative type system for \Calc
    (\Cref{sec:declarative}).
  We start with a succinct declarative type system that uses colours
  to denote directions of information flow for polytypes, and then
  modularly extend it to incorporate skeleton inference.
  \item We give an algorithmic type system for \Calc and prove its
    soundness and completeness with respect to the declarative system
    (\Cref{sec:algorithmic}).
  We only introduce unification variables for monotypes.
  Polymorphic instantiation is resolved locally via a novel look
  judgement.
  \item We provide a full implementation of the algorithm of \Calc
  with extensions, and a faithful encoding of the declarative type
  system of \Calc in Rocq (\Cref{sec:implementation}).
  We demonstrate the scalability of \Calc by extending the
  implementation to infer modal effect types~\citep{TangWDHLL25}.
\end{itemize}
\Cref{sec:related-work} discusses related work and future work.

\FloatBarrier
\section{Overview}
\label{sec:overview}

In this section, we give a high-level overview of \Calc.
We begin with a brief recap of conventional bidirectional type
systems.
Then we explain the key technical ideas of \Calc one by one, including
skeletons and ghosts, colours and modes, polymorphic subtyping with
tags, and freezing.
Finally we provide more examples and discuss the limitations of \Calc.

\subsection{Conventional Bidirectional Type Systems}
\label{sec:conventional-bidirectional}

Conventional bidirectional type systems split typing into inference
mode $\TI$, where we infer the type of a term, and checking mode
$\TC$, where we check a term against a given type.
A function application rule typically has the following shape.
\[\small
  \inferrule*
  {
    \stypi{\Gamma}{M}{A} \\
    \text{inst}(A) = A_1 \to B \\
    \stypc{\Gamma}{N}{A_1}
  }
  {\stypi{\Gamma}{M\;N}{B}}
\]
The application itself is in inference mode.
We first infer the type $A$ of the function, then instantiate its type
to a function type $A_1\to B$, and finally check the argument against
the argument type $A_1$.
Thus information flows from functions to arguments.
Such information flow naturally solves challenge (2) in
\Cref{sec:introduction}, assuming the argument type is known.
In the motivating example, if we are given the type
$(\forall a .a\to a)\to \Pair{\Int}{\Bool}$,
type checking the first argument proceeds as follows.
\[\small
  \inferrule*
  {
    \stypc{\Gamma, f : \forall a . a \to a}
      {(f\;42, f\;\True)}
      {\Pair{\Int}{\Bool}}
  }
  {
    \stypc{\Gamma}{\lambda f . (f\;42, f\;\True)}
      {(\forall a . a \to a) \to \Pair{\Int}{\Bool}}
  }
\]
The checking rule of lambda abstraction assigns the known polytype
$\forall a . a \to a$ to the parameter $f$.

The missing step is the polymorphic instantiation in challenge (1).
Many existing bidirectional type systems restrict attention to
higher-rank
polymorphism~\citep{DunfieldP04,DunfieldK13,CuiJO23,ZhaoO22,JiangCO25}
and only allow instantiation with monotypes.
We cannot simply extend such systems with arbitrary implicit
polymorphic instantiation as this is an undecidable
problem~\citep{chrzaszcz1998poly,tiuryn1996subtyping}.
However, in our example the second argument $\churchIds$ provides the
information to instantiate $\churchMap$ with: the polytype $\forall a
. a \to a$.
Thus, in order to address challenge (1), we naturally wish
information to flow from arguments to functions.
\Fresco enables such information flow via skeletons and ghosts.

\subsection{Skeletons and Ghosts}
\label{sec:overview-skeletons}

In order to perform polymorphic instantiation, we must collect
information from the argument $\churchIds$ of type
$\ChurchList\;(\forall a . a \to a)$, which expands to $\forall b . b
\to ((\forall a . a \to a) \to b\to b) \to b$.
Directly using the argument type is, however, not always sufficient.
Even though this works in the motivating example, in general, further
implicit instantiation or generalisation may occur on the argument,
before the argument type of the function can be matched against.
For instance, in $\dec{foo}\;\churchIds$ where $\dec{foo}: (A\to
((\forall a . a \to a) \to A \to A) \to A) \to B$, the type of
$\churchIds$ is instantiated with $A$.
As a result, we should make sure that the information we collect from
the argument is compatible with the types of the argument after any
instantiation or type abstraction.

\Fresco uses \emph{skeletons} to represent the information collected
from arguments.
Skeletons extend types with \emph{ghosts}, including both ordinary
ghosts $\Hole$, which represent unknown types, and universal ghosts
$\TopHole$, which represent unknown (possibly empty) sequences of
universal quantifiers.
For instance, we infer the following skeleton for $\churchIds$.
\[
  \TopHole . \Hole \to ((\forall a.a\to a)\to\Hole\to\Hole) \to \Hole
\]
This skeleton is consistent with any possible type that $\churchIds$
may have.
Note that $\TopHole$ is not a binder itself, and all instances of
ghosts in a skeleton are not related.
The four appearances of $\Hole$ can match any other types we use to
instantiate $\churchIds$.
The top-level $\TopHole$ can match any number of universal
quantifiers introduced by type abstraction on $\churchIds$.
As a result, in the example $\dec{foo}\;\churchIds$ defined above, the
skeleton of $\churchIds$ is consistent with the argument type of
$\dec{foo}$.
Meanwhile, the skeleton still shows that list elements have type
$\forall a.a\to a$.

As another example, the skeleton of the first argument $\lambda f .
(f\;42, f\;\True)$ to $\churchMap$ is
    ${\TopHole . \Hole \to \TopHole.\Pair{\Hole}{\Hole}}$.
It tells us that the term should have a possibly polymorphic
function type with an unknown argument type and a possibly polymorphic
product return type.
Again, this skeleton is consistent with any possible type for a lambda
abstraction of this shape.
In particular, it does not commit the type of $f$ to be any specific
type, but leaves it as a ghost $\Hole$.
Later, this information can flow back from the argument type of
$\churchMap$ to the lambda abstraction and resolve challenge (2), as
illustrated in \Cref{sec:conventional-bidirectional}.

Skeletons only carry polymorphic information that we are certain
about.
For instance, the skeleton of $\dec{id}$ cannot be
$\forall\alpha.\alpha\to\alpha$, because the quantifier
$\forall\alpha$ may be instantiated.
On the other hand, the quantifier $\forall\alpha$ in the skeleton of
$\churchIds$ is allowed, because it is in a guarded position (under
function arrows) which cannot be instantiated.

Skeleton inference rules of \Fresco correspond one-to-one with the
syntax-directed typing rules of \Fresco with systematic minimal
changes (\Cref{sec:skeleton-inference}).
We believe that skeleton inference smoothly scales to support other
typing features.
We provide some evidence for this belief by extending our
implementation of \Fresco with effects and handlers along with a modal
effect type system~\citep{TangWDHLL25,TangL26}.

\subsection{Colours}
\label{sec:overview-colours}

Information provided by skeletons of arguments is known.
When typing the function, we want to synthesise its type guided by
such information inherited from the arguments.
We use colours, following coloured local type inference
(CLTI)~\citep{OderskyZZ01}, to distinguish between synthesised and
inherited parts of a type.
Unlike in CLTI~\citep{OderskyZZ01}, we allow monotypes to be freely
guessed as they can be solved efficiently by unification.
As a result, we need only ascribe colours to universal quantifiers,
which cannot be guessed.
For instance, we can fill in the ghosts of the skeleton of
$\churchIds$ to obtain a coloured type (which we call a
\emph{bidirectional} type) as follows for the function $\churchMap$.
\[
  \syn{\forall b} . b \to ((\inh{\forall a}.a\to a)\to b\to b) \to b
\]
The quantifier $\inh{\forall a}$ is inherited ({\color{cblue}{blue}})
since it is provided by the skeleton of the argument $\churchIds$.
The quantifier $\syn{\forall b}$ is synthesised
({\color{cred}{red}}) since it is inferred from the function
$\churchMap$; the skeleton of $\churchIds$ only provides a universal
ghost $\TopHole$ in this position.
Monotypes such as function arrows and type variables are always
uncoloured.
When writing a monotype with colours, we mean the same type as the
uncoloured one. For instance, $\syn{\Int}$ is the same as $\Int$.

We colour type meta variables as well as quantifiers.
We let $A$ range over uncoloured types (quantifiers are uncoloured),
$\syn{A}$ range over synthesised types (quantifiers are synthesised),
$\inh{A}$ range over inherited types (quantifiers are inherited), and
$\bid{A}$ range over bidirectional types (quantifiers are either
synthesised or inherited).
Type meta variable names are \emph{colour blind} in the sense that
(e.g. in typing rules) $A$, $\syn{A}$, $\inh{A}$, and $\bid{A}$ all
implicitly denote the same type up to colouring of quantifiers.
We also colour skeletons and have the same conventions for $P$,
$\syn{P}$, $\inh{P}$ and $\bid{P}$.

We define the above operation of filling in the ghosts %
as a skeleton refinement relation $\inh{P}\fillsyn\bid{A}$.
Here, $\inh{P}$ is an inherited skeleton, and $\bid{A}$ is a
bidirectional type with inherited information from $\inh{P}$ and new
synthesised information.
For instance, the above skeleton refinement is written as:
\[
  \TopHole . \Hole \to ((\inh{\forall a}.a\to a)\to\Hole\to\Hole) \to \Hole
  \fillsyn
  \syn{\forall b} . b \to ((\inh{\forall a}.a\to a)\to b\to b) \to b
\]

The refinement relation characterises what other types a skeleton
represents.
We cannot fill in ghosts with arbitrary types.
In $\inh{P}\fillsyn\bid{A}$, we disallow use of variables introduced
by inherited quantifiers in $\inh{P}$ to fill in ghosts.
For instance, $\inh{\forall a} .\Hole \to a \fillsyn \inh{\forall a}
.a\to a$ is invalid, because we use the bound variable $a$ to fill in
a ghost, even though the result type is well-scoped.
This restriction guarantees that when a quantifier is known, all
appearances of the bound variable must also be known, since these
variable appearances also describes the shape of the polytype.

\subsection{Modes}
\label{sec:overview-modes}

In order to type check the motivating example, we need to utilise the
information from the second argument.
However, a standard application rule such as the one in
\Cref{sec:conventional-bidirectional} only takes one argument into
account.
Some previous work~\citep{PierceT00,SerranoHJV20} gives the
application spine $M\;\ol{N}$ a special typing rule.
However, it is not stable under term transformations such as adding
a dummy lambda abstraction, e.g., $(\lambda x.M)\;42\;\ol{N}$.
We adopt a more general approach to dealing with argument terms.
As shown in \Cref{sec:conventional-bidirectional}, bidirectional
typing typically has two modes $\TI$ and $\TC$.
Inspired by counters in Contextual Typing~\citep{XueO24}, we add a
function mode form $\FN\M$ (where $\M$ ranges over modes),
for terms in function position.
Our modes are primarily used to track local program structure, unlike
traditional bidirectional typing where modes fully determine the
direction of information flow.
For instance, the typing judgement for $\churchMap$ in our running
example has the form $\typp[\FN\FN\TI]{\Gamma}{\churchMap}{\bid{A}}$
for some type $\bid{A}$.
The mode $\FN\FN\TI$ merely tells us that it is applied to two
arguments, and that the whole application is in an inference position
with no input type.
Information flow is determined by colours in the bidirectional type
$\bid{A}$.

\subsection{Flowing Back and Forth}
\label{sec:overview-application}

With all the above ideas (skeletons, ghosts, colours, and modes), we
now show the key rule of \Calc, the typing rule for function
application.
This rule enables information to flow back and forth between
functions and arguments.
\[\small
  \inferrule*
  {
    \refn[1][\pati{\Gamma}{N}{\syn{P}}] \\
    \refn[2][\Gamma \vdash \inh{P} \fillsyn \bid{A}] \\
    \refn[3][\typp[\FN\M]{\Gamma}{M}{\bid{A} \to \bid{B}}] \\
    \refn[4][\typc{\Gamma}{N}{\inh{A}}]
  }
  {\typp[\M]{\Gamma}{M\;N}{\bid{B}}}
\]
Though the rule is declarative, it is easiest to explain via an
algorithmic intuition.
Skeleton inference \refn[1] first infers a synthesised skeleton
$\syn{P}$ for the argument $N$, which carries the polytype information
gleaned from $N$ itself.
Since $P$ is now known, skeleton refinement \refn[2] refines inherited
$\inh{P}$ to a bidirectional type $\bid{A}$, which is used as the
argument type of the function $M$ in the typing judgement \refn[3].
Inherited quantifiers in $\bid{A}$ come from the argument $N$ in
\refn[1], while synthesised quantifiers in $\bid{A}$ are inferred from
the function $M$ in \refn[3].
Finally, $A$ is fully known and used to type check the argument $N$ in
\refn[4].
In summary, information first flows from the argument to the function,
and then flows back from the function to the argument.
This differs from the application rule in
\Cref{sec:conventional-bidirectional}, where information only flows
from function to argument.

It is worth noting that skeleton inference \refn[1] uses checking mode
$\TC$.
This is because $N$ is in a checking position where potential
instantiation or type abstraction may happen.
Skeleton inference takes this into account and infers a skeleton that
is consistent with any possible type of $N$.

As an example, we show how the running example is typed using the
application rule.
With $P_{\texttt{ids}} = \TopHole . \Hole \to ((\forall a.a\to
a)\to\Hole\to\Hole) \to \Hole$
and $\syn{\ChurchList}\;(\inh{\forall a}.a\to a) = \syn{\forall b} . b
\to ((\inh{\forall a}.a\to a)\to b\to b) \to b$, the typing of the
outer application goes as follows.
\[\small
  \inferrule*
  {
    \pati{\Gamma}{\churchIds}{\hl{\syn{P_{\mathit{ids}}}}} \\
    \Gamma \vdash \hl{\inh{P_{\mathit{ids}}}} \fillsyn
      \hl{\syn{\ChurchList}\;(\inh{\forall a}.a\to a)} \\\\
    \typp[\FN\TI]{\Gamma}{\churchMap\;\text{$(\lambda f . (f\;42, f\;\True))$}}
      {\hl{\syn{\ChurchList}\;(\inh{\forall a}.a\to a)}
        \to\syn{\ChurchList}\;(\Pair{\Int}{\Bool})} \\
    \typc{\Gamma}{\churchIds}{\hl{\inh{\ChurchList}\;(\inh{\forall a}.a\to a)}}
  }
  {\typp[\text{$\TI$}]{\Gamma}
    {\churchMap\;\text{$(\lambda f . (f\;42, f\;\True))$}\;\churchIds}
    {\syn{\ChurchList}\;(\Pair{\Int}{\Bool})}}
\]
In each premise, we highlight the part that demonstrates how
information flows from the argument to the function and back.
We infer the skeleton $\syn{P_{\texttt{ids}}}$ of $\churchIds$, refine
it to the bidirectional type $\syn{\ChurchList}\;(\inh{\forall a}.a\to
a)$, use this type as the argument type for the inner application at
mode $\FN\TI$, and finally type check $\churchIds$.
The typing of the inner application goes as follows.
\[\small
  \inferrule*
  {
    \pati{\Gamma}{\text{$(\lambda f . (f\;42, f\;\True))$}}
      {\hl{\syn{\TopHole . \Hole \to \TopHole.\Pair{\Hole}{\Hole}}}} \\
    \Gamma \vdash
      \hl{\inh{\TopHole . \Hole \to \TopHole.\Pair{\Hole}{\Hole}}}
      \fillsyn
      \hl{(\syn{\forall c .} c \to c) \to \Pair{\Int}{\Bool}} \\\\
    \typp[\FN\FN\TI]{\Gamma}{\churchMap}
      {\hl{((\syn{\forall c .} c \to c) \to \Pair{\Int}{\Bool})}
        \to\syn{\ChurchList}\;(\inh{\forall a}.a\to a)
        \to\syn{\ChurchList\;(\Pair{\Int}{\Bool})}} \\
    \typc{\Gamma}{\lambda f . (f\;42, f\;\True)}
      {\hl{(\inh{\forall c .} c \to c) \to \Pair{\Int}{\Bool}}}
  }
  {\typp[\FN\TI]{\Gamma}{\churchMap\;\text{$(\lambda f . (f\;42, f\;\True))$}}
    {\syn{\ChurchList}\;(\inh{\forall a}.a\to a)
      \to\syn{\ChurchList\;(\Pair{\Int}{\Bool})}}} \\
\]
In this case the skeleton $\syn{\TopHole . \Hole \to
\TopHole.\Pair{\Hole}{\Hole}}$ of the argument $\lambda f . (f\;42,
f\;\True)$ does not really provide any polymorphic information.
The function $\churchMap$ infers the quantifier $\syn{\forall c}$,
which is then used as inherited information to type check the argument.

\subsection{Polymorphic Subtyping with Tags}
\label{sec:overview-subtyping}

We have seen how information flows back and forth.
Now we focus on how the polymorphic instantiation of $\churchMap$ is
performed in its typing judgement:
\[\small
\typp[\FN\FN\TI]{\Gamma}{\churchMap}
  {
    ((\syn{\forall c .} c \to c) \to \Pair{\Int}{\Bool})
    \to  \syn{\ChurchList}\;(\inh{\forall c} . c \to c)
    \to \syn{\ChurchList\;(\Pair{\Int}{\Bool})}
  }
\]
The second argument type carries the inherited quantifier
$\inh{\forall c}$ from the skeleton of $\churchIds$.
Recall that the type of $\churchMap$ is
  $\forall a\, b . (a \to b) \to \ChurchList\;a \to \ChurchList\;b$.
We want to instantiate the type of $\churchMap$ to the bidirectional
type above.
As usual, we combine instantiation together with type abstraction into
a subtyping relation.
For the example here, the subtyping is:
\[\small\bl
\typsub[\FN\FN\TI]{\Gamma}{
  \syn{\forall a\, b} . (a \to b) \to \syn{\ChurchList}\;a \to \syn{\ChurchList}\;b
}{
  \\ \qquad\quad\ \
    ((\syn{\forall c .} c \to c) \to \Pair{\Int}{\Bool})
    \to  \syn{\ChurchList}\;(\inh{\forall c} . c \to c)
    \to \syn{\ChurchList}\;(\Pair{\Int}{\Bool})
}
\el\]

Intuitively, this judgement upcasts the LHS to the RHS guided by
inherited information on the RHS.
We must instantiate $a$ to a polymorphic type $\forall c . c \to c$ as
determined by the inherited information from the second argument.
In a traditional bidirectional type system such as that of
\citet{DunfieldK13}, the declarative subtyping rule guesses a
monomorphic type for instantiation, which is solved by unification.
In \Calc, though we can still guess monotypes, we cannot guess
polytypes arbitrarily.
Otherwise, for instance, $\dec{single}\;\dec{id}$ would have
possible types $\List\;(\forall a.a\to a)$ and $\List\;(A\to A)$ with
no best choice.
Instead, we want to guess polytypes only based on the inherited
information we know.
However, it is challenging to distil the idea of ``never guess
polymorphism out of thin air'' in a declarative way.
This is one of the main reasons that many type systems for FCP lack
declarative specification and completeness results.
For instance, QuickLook~\citep{SerranoHJV20}, which also collects
information from arguments to guide polymorphic instantiation of
functions, only has an algorithmic specification of instantiation.

We propose a novel approach to characterise the idea of ``never guess
polymorphism out of thin air'' in a declarative way, inspired by work
on provenance tracking~\citep{BunemanKT01} and residuals in lambda
calculus~\citep{Barendregt85}.
Within a polytype we guess for instantiation, we tag each quantifier
with a unique tag, and track whether each quantifier is determined by
inherited information via its tag.
For instance, in the subtyping judgement above, for the type
variables $a$ and $b$ we guess
  $\syn{\varphi_a} = \syn{\foralltag{\epsilon} c} . c \to c$
  and
  $\syn{\varphi_b} = \Pair{\Int}{\Bool}$.
The tag $\epsilon$ is fresh.
After substituting $\varphi_a$ for $a$, the second argument type of
the function type becomes
  $\syn{\ChurchList}\;(\syn{\foralltag{\epsilon} c} . c \to c)$, which
  matches the argument type on the RHS:
\[
  \typconsis[\epsilon]{\Gamma}
    {\syn{\ChurchList}\;(\syn{\foralltag{\epsilon} c} . c \to c)}
    {\syn{\ChurchList}\;(\inh{\forall c .} c \to c)}.
\]
This \emph{consistency} judgement determines the tag $\epsilon$ via
the inherited quantifier $\forall c$ on the RHS.
The judgement includes a tag context (here the singleton $\epsilon$).

Tags share information about whether a guessed quantifier is
determined across different appearances of the same quantifier.
A guessed quantifier is determined as long as one of its appearances
is determined by inherited information.
For instance, in the working example, $\varphi_a$ appears twice in the
instantiated type, but it is determined fully by its second appearance
in the second argument type $\syn{\ChurchList}\;(\syn{\foralltag{\epsilon} c}
. c \to c)$.
This improves over prior work such as CLTI~\citep{OderskyZZ01} and
Boxy types~\citep{VytiniotisWJ06}.
These approaches also support mixed information flow through ideas
similar to colours and skeletons.
However, they have a stricter restriction on polymorphic
instantiation, requiring all appearances of the guessed instantiation
to be determined, rather than just one.

We now present a negative example where subtyping fails as it would
require guessing polymorphism out of thin air.
Consider $\dec{single}\;\dec{id}$.
We may have the invalid subtyping judgement:
\[\small
\typsub[\FN\TI]{\Gamma}{
  \syn{\forall a} . a \to \List\;a
}{
  (\syn{\forall b} . b\to b) \to \List\;(\syn{\forall b}. b\to b)
}
\]
Guessing $\syn{\varphi_a} = \syn{\foralltag{\epsilon} b} . b\to b$ for
the instantiation, we have the subtyping judgement:
\[\small
\typsub[\FN\TI]{\Gamma}{
  (\syn{\foralltag{\epsilon} b} . b\to b) \to \List\;(\syn{\foralltag{\epsilon} b} . b\to b)
}{
  (\syn{\forall b} . b\to b) \to \List\;(\syn{\forall b}. b\to b)
}
\]
We cannot determine the tag $\epsilon$ via this judgement, as both
appearances of $\syn{\foralltag{\epsilon} b}$ correspond to
synthesised $\syn{\forall b}$ on the RHS.
As a result, in \Calc, $\List\;(\forall a. a\to a)$ is not a valid
type for $\dec{single}\;\dec{id}$.
This solves the no best choice problem.

Our algorithm for subtyping does not guess types for instantiation.
Instead, for an instantiation, it uses a look judgement to inspect the
RHS for the quantified variable.
Look is separated into constraint collection and constraint solving.
In the motivating example, constraint collection for $a$ extracts from
the correspondence between $\ChurchList\;a$ and
$\ChurchList\;(\inh{\forall c .} c \to c)$ that the type for $a$
should be consistent with $\forall c .c \to c$.
Constraint solving then determines the instantiation for $a$ as
$\forall c .c \to c$.
One advantage of look is that it is local: it only happens in
subtyping for solving instantiation, and it does not introduce any
unification variable to the context.
As a result, our algorithm only does standard global unification for
monotypes.

\subsection{Freezing}
\label{sec:overview-freezing}

The motivating example does not require freezing.
The skeleton of $\churchIds$ already provides enough information.
Freezing is useful when argument skeletons do not provide enough
information.
For instance, in $\dec{single}\;\dec{id}$, \Calc does not infer the
type $\List\;(\forall a . a \to a)$.
The reason is that the skeleton of $\dec{id}$ is $\TopHole .  \Hole
\to \Hole$.
We cannot give $\dec{id}$ a more concrete skeleton such as $\forall
a.a\to a$, as a skeleton should be consistent with any possible type
of the term (\Cref{sec:overview-skeletons}).

Freezing enables a programmer to fine-tune the information flowing
from arguments to functions.
Freezing a term disables any potential instantiation or type
abstraction on it, regardless of the context.
We write $\Freeze{M}$ for freezing a term $M$.
For instance, $\Freeze{\dec{id}}$ has skeleton $\forall a . a \to a$.
As a result, \Calc infers the type $\List\;(\forall a.a\to a)$ for
$\dec{single}\;\Freeze{\dec{id}}$ via the following derivation.
\[\small
  \inferrule*
  {
    \pati{\Gamma}{\Freeze{\dec{id}}}{\hl{\syn{\forall b} . b \to b}} \\
    \Gamma \vdash
      \hl{\inh{\forall b .} b \to b}
      \fillsyn
      \hl{\inh{\forall b .} b \to b} \\\\
    \typp[\FN\TI]{\Gamma}{\dec{single}}
      {\hl{(\inh{\forall b .} b \to b)} \to
        \List\;(\syn{\forall b .} b \to b)} \\
    \typc{\Gamma}{\Freeze{\dec{id}}}{\hl{\inh{\forall b} . b \to b}}
  }
  {
    \typi{\Gamma}{\dec{single}\;\Freeze{\dec{id}}}
      {\List\;(\syn{\forall b .} b \to b)}
  }
\]
The polymorphic instantiation of the function $\dec{single}$ succeeds
because the first $\forall b$ is inherited.

Though freezing an argument provides more information for the
function, it is not always desirable.
For instance, $\dec{run}\;\Freeze{\dec{id}}$ where
$\dec{run}:(\Int\to\Int)\to\Int$ is ill-typed, because $\dec{run}$
expects a function type $\Int\to\Int$, yet $\Freeze{\dec{id}}$
disables instantiation on $\dec{id}$.

We give one more example of freezing based on a variant of our
motivating example.
With a standard application function $\dec{app} : \forall a\, b . (a
\to b) \to a \to b$, the following application is ill-typed.
\[
  \dec{app}\;(\lambda f . (f\;42, f\;\True))\;{\dec{id}}
\]
The argument $\dec{id}$ does not provide enough information to
instantiate $\dec{app}$ with a polytype $\forall a . a\to a$, unlike
$\churchIds$ in the motivating example.
We can make it well-typed by freezing $\dec{id}$ as follows.
\[
  \dec{app}\;(\lambda f . (f\;42, f\;\True))\;\Freeze{\dec{id}}
\]
Now $\dec{app}$ instantiates with $\forall a . a \to a$, which then
flows back to the lambda abstraction and assigns a polytype to $f$.

\subsection{Features and Design Choices}

We discuss some important features and design choices of \Fresco with
examples.

\paragraph{Order irrelevance.}
\Fresco is not sensitive to argument order.
This observation is exemplified by our motivating example, where the
second argument $\churchIds$ provides the information to instantiate
the function $\churchMap$, and this information flows back to the
first argument $\lambda f . (f\;42, f\;\True)$.
As another simple example, consider two functions $\dec{choose} :
\forall a .  a \to a \to a$ and $\dec{auto} : (\forall b . b\to b) \to
(\forall b .b\to b)$. \Calc infers the same type for both
$\dec{choose}\;\dec{auto}\;\dec{id}$ and
$\dec{choose}\;\dec{id}\;\dec{auto}$.
In both cases, $\dec{auto}$ guides the instantiation of $\dec{choose}$
with $(\forall b . b\to b) \to (\forall b .b\to b)$, which then flows
back to $\dec{id}$ and instantiates $\dec{id}$ with $\forall b . b\to
b$.

\paragraph{Polymorphic instantiation guided by result type.}
\Cref{sec:overview-subtyping} illustrates how argument types guide
polymorphic instantiation of functions.
Inherited result types can also guide polymorphic instantiation.
For instance, for the type checking judgement
$\typp[\TC]{\Gamma}{\dec{nil}}{\List\;(\inh{\forall b}.b\to b)}$
where $\dec{nil}:\forall a.\List\;a$,
its subtyping judgement is
$\typsub[\TC]{\Gamma}{\syn{\forall a} . \List\;a}{\List\;(\inh{\forall b}.b\to b)}$.
We guess $\syn{\foralltag{\epsilon} b} . b\to b$ for instantiation,
which is determined by the consistency
$\typconsis[\epsilon]{\Gamma}{\syn{\foralltag{\epsilon} b} . b\to
b}{\inh{\forall b}.b\to b}$.

\paragraph{Flowing back and forth within one argument}
We have seen how information flows from one argument to the function,
and then flows back to another argument.
\Calc also allows information to flow back to the same argument.
For instance, consider the application $\dec{foo}\;\dec{bar}$
where $\dec{foo} : \forall a . ((\forall b . b \to b) \to a) \to a$
and $\dec{bar} : \forall c . c \to (\forall d . d\to d)$.
The argument guides the function to instantiate $a$ with $\forall d .
d\to d$, while the function guides the argument to instantiate $c$
with $\forall b . b \to b$.

\paragraph{Further instantiation}
As one would expect, \Calc allows types used for instantiation to be
further instantiated.
For instance, given functions $\dec{head} : \forall a . \List\;a \to
a$ and $\dec{ids} : \List\;(\forall a . a \to a)$, \Calc infers the
type $\Int$ for $\dec{head}\;\dec{ids}\;42$.
The first application $\dec{head}\;\dec{ids}$ instantiates
$\dec{head}$ with $\forall a . a \to a$, which is the result type
that is further instantiated with $\Int$ to match $42$.

\paragraph{Shallow subtyping}
\Calc only considers shallow subtyping, i.e., instantiation and type
abstraction apply only to top-level types of terms.
This is a common design choice for FCP type systems, as it enables
more polymorphism to be inferred.
For instance, in our motivating example, the skeleton of $\churchIds$
only contains the quantifier because it is in a guarded position
(inside a function arrow) and thus can never be instantiated.
In contrast, the skeleton of $\dec{id}$ is $\TopHole . \Hole \to
\Hole$ without the top-level quantifier, because it can be
instantiated.
Modes play a central role in subtyping, determining when subtyping is
allowed.
For instance, in the subtyping judgement
\[
\typsub[\FN\TC]{\Gamma}{\Int \to \forall a . a\to a}{\Int \to \Int \to \Int}
\]
we can still instantiate $\forall a$ without breaking the shallow
subtyping restriction as the function mode $\FN\TC$ specifies that
the term is in function position, and we can instantiate the whole
function application term of type $\forall a.a\to a$ with $\Int$.

\paragraph{Value restriction}
\Calc adopts the value restriction~\citep{SML,Wright95}.
Type abstraction only applies to syntactic values.
For instance, given $\dec{poly} : (\forall a.a\to a) \to \Int$, the
term $\dec{poly}\;(\lambda x. x)$ is well-typed as information flowing
from the function to the argument introduces type abstraction on the
argument $\lambda x. x$.
In contrast, $\dec{poly}\;(\dec{id}\;\dec{id})$ is ill-typed as we
cannot generalise the application $\dec{id}\;\dec{id}$ because it is
not a value.
To make it well-typed, we can either $\eta$-expand the argument to a
value as $\dec{poly}\;(\lambda x. \dec{id}\;\dec{id}\;x)$, or freeze
the second $\dec{id}$ as $\dec{poly}\;(\dec{id}\;\Freeze{\dec{id}})$.
Freezing allows us to directly infer the type $\forall a.a\to a$ for
the argument $\dec{id}\;\Freeze{\dec{id}}$ without a type abstraction.
The value restriction is necessary for our extension of \Calc with
effects and inference for modal effect types in
\Cref{sec:implementation}.

\subsection{Limitations}
\label{sec:overview-limitations}

We discuss some limitations of \Calc with examples.

\paragraph{Failures}
Consider the subtyping judgement $\typsub[\FN\TC]{\Gamma}{\syn{\forall
a}. \tunit\to a}{\tunit\to\inh{\forall b}. b \to b}$.
The LHS type is uninhabited.
Unfortunately, this subtyping judgement is invalid in \Calc.
Suppose we guess $\syn{\foralltag{\epsilon} c}.c\to c$ as the
instantiation of $a$.
After instantiation we have
$\typsub[\TC]{\Gamma}{\syn{\foralltag{\epsilon} c}.  c\to
  c}{\inh{\forall b}. b \to b}$.
The quantifier $\foralltag{\epsilon} c$ is not determined by
$\inh{\forall b}$ because we can still further instantiate it.
However, for the original subtyping judgement, the instantiation with
$\syn{\foralltag{\epsilon} c}.c\to c$ is obviously the simplest
solution.
Extending \Calc to support such subtyping judgements without violating
the ``never guess polymorphism out of thin air'' principle is an
interesting direction for future work.

\paragraph{Ghosts are nameless.}

In \Calc, ghosts are nameless.
Being nameless keeps skeletons simple, but also loses the connection
between different occurrences of ghosts.
For instance, consider the application
$\dec{compose}\;\dec{poly}\;\dec{head}$ with type signatures:
\[\small
  \dec{compose} : \forall a\, b\, c . (b \to c) \to (a \to b) \to a \to c
  \quad\
  \dec{head} : \forall d . \List\;d \to d
  \quad\
  \dec{poly} : (\forall e . e\to e) \to \Int
\]
The skeleton of $\dec{poly}$ is its type $(\forall e.e\to e)\to\Int$.
The skeleton of $\dec{head}$ is $\TopHole . \List\;\Hole \to \Hole$,
which fails to capture the connection between the argument and result
types of $\dec{head}$.
The skeleton of $\dec{poly}$ guides instantiation of $b$ with $\forall
e . e\to e$.
Type checking $\dec{head}$ would require instantiating $a$ with the
polytype $\List\;(\forall e . e\to e)$, which is not possible as the
inherited information is too coarse.

In other systems such as QuickLook, this program is well-typed.
QuickLook directly uses unification variables when specifying the
type inference algorithm, which preserves the connection between
argument and result types of $\dec{head}$.
We intentionally omit unification variables from the declarative
system of \Calc for simplicity and predictability.
In order to permit such examples, a promising extension to explore is
named ghosts.
We defer this exploration to future work.

\paragraph{Only flowing back and forth once}

\Calc only allows information to flow back and forth once, first
flowing from the argument to the function, and then flowing back from
the function to the argument.
It can be helpful for information to flow back and forth multiple
times between function and argument.
For instance, consider the application $\dec{f}\;\dec{g}$ with two
variables:
\[
f : \forall c\, d . (c \to c \to \List\;d) \to \Int \qquad
g : \forall b .(\forall a . a\to a) \to b \to \List\;b
\]
The skeleton of $\dec{g}$ is $\TopHole . (\forall a . a\to a) \to
\Hole \to \Hole$, which guides $f$ to instantiate $c$ with $\forall a
. a\to a$.
This information then flows back to $g$ and guides it to instantiate
$b$ with $\forall a . a \to a$.
However, in order to make it well-typed, we should instantiate $d$
with $\forall a . a\to a$ as well, which requires information flow
from $g$ to $f$ again.
We intentionally do not allow such recurring information flow in \Calc
for efficiency and predictability reasons.
We defer exploring such an extension to future work.

\FloatBarrier
\section{Declarative Type System}
\label{sec:declarative}

In this section, we give the declarative presentation of \Calc with
freezing, skeletons, colours, and ghosts.
For clarity, we first present the syntax of types and terms, typing,
and subtyping for \Calc, and then introduce skeletons and skeleton
inference.

\subsection{Syntax, Colours, and Modes}

The syntax of types, terms, contexts, and modes for \Calc is as
follows.
  \[\small
  \ba[t]{@{}=l@{\mkern9mu}+r@{~}+c@{~}+l@{}}
  \slab{Types}    &A,B  &::= & \tunit \mid \alpha
                            \mid A\to B
                            \mid \forall\alpha . A
                            \\
  \slab{Synthesised types}   &\syn{A}, \syn{B} &::= &
    \tunit \mid {\alpha} \mid \syn{A} \to \syn{B}
    \mid \syn{\forall\alpha .} \syn{A}
    \\
  \slab{Inherited types} &\inh{A}, \inh{B} &::= &
    \tunit \mid {\alpha} \mid \inh{A} \to \inh{B}
    \mid \inh{\forall\alpha .} \inh{A}
    \\
  \slab{Bidirectional types}    &\bid{A},\bid{B}  &::= &
    \tunit \mid {\alpha} \mid \bid{A} \to \bid{B}
    \mid \inh{\forall\alpha .} \bid{A}
    \mid \syn{\forall\alpha .} \bid{A}
    \\
  \slab{Modes} &\M &::= & \TI \mid \TC \mid \FN\M \\
  \ea
  \quad
  \ba[t]{@{}=l@{\mkern3mu}+r@{~}+c@{~}+l@{}}
  \slab{Terms}          &M,N  &::= &
    \eunit \mid x
    \mid {\lambda x.M}
    \mid \lambda x^A.M
    \mid {M:A} \\
    & &\mid &
    {\Freeze{M}}
    \mid M\;N
    \\
  \slab{Values}         &V  &::= &
    \eunit \mid x
    \mid \lambda x.M
    \mid \lambda x^A.M
    \mkern1mu\mid {V : A}
    \\
  \slab{Contexts} &\Gamma &::=&
    \cdot \mid \Gamma, \alpha
    \mid \Gamma, x : A
    \\
  \ea\]

Types include a unit type ($\TUnit$), type variables ($\alpha$),
function types ($A\to B$), and universal types ($\forall\alpha.A$).
As usual, we have standard well-scopedness ($\Gamma\vdash A$) and
$\alpha$-equivalence ($\Gamma\vdash A\equiv B$) on types, including
coloured types, as well as terms.
We adopt the Barendregt convention that all variables are
distinct~\citep{Barendregt85}.
We say that a type is guarded if it has no top-level quantifiers.

Modes, as introduced in \Cref{sec:overview-modes}, track local program
structure, indicating whether a term is in an inference position
($\TI$), a checking position ($\TC$), or a function position ($\FN\M$)
with function body at mode $\M$.
Function modes specify whether the argument type comes from an
argument term.

Colours, as introduced in \Cref{sec:overview-colours}, track the
direction of information flow for polytypes.
We write plain types in black, synthesised types in $\color{cred}{\text{red}}$, inherited
types in $\color{cblue}{\text{blue}}$, and bidirectional types in \gray{grey}.
We only care about whether a quantifier is synthesised or inherited;
monotypes are always black.
A bidirectional type $\bid{A}$ tracks whether each quantifier is
synthesised ($\syn{\forall\alpha}$) or inherited
($\inh{\forall\alpha}$).
Types in term annotations and contexts remain black.
Consequently, programmers never write coloured types in annotations,
and never see variable bindings with coloured types in contexts.

The syntax of terms and contexts is mostly standard.
We include a freezing operator $\Freeze{M}$ inspired by
\FreezeML~\citep{EmrichLSCC20}, which allows us to freeze the type
synthesised for $M$ in the current context, and thus treat it as
inherited.
We have a syntactic category of values $V$ for value
restriction~\citep{SML,Wright95}.

\FloatBarrier
\subsection{Typing Rules}
\label{sec:typing}

\Cref{fig:decl-typing} gives the declarative typing rules
for \Calc.
We highlight judgements we have not yet defined in $\hl{\text{light
grey}}$ background.
The typing judgement has the form $\typp[\M]{\Gamma}{M}{\bid{A}}$,
which states that under context $\Gamma$ in mode $\M$ the term $M$ has
type $\bid{A}$.
As usual, we require standard well-scopedness of $M$ and $\bid{A}$
under $\Gamma$ as well-formedness conditions for the typing judgement.
Moreover, we require compatibility between coloured types and modes
$\skmod[\TY]{\bid{A}}{\M}$.
The relation $\skmod[\TY]{\bid{A}}{\M}$ is defined as
${\skmod[\TY]{\syn{A}}{\TI}}$,
${\skmod[\TY]{\inh{A}}{\TC}}$,
and
${\skmod[\TY]{\bid{A}\to\bid{B}}{\FN\M}}$ if
$\skmod[\TY]{\bid{B}}{\M}$.

This compatibility relation demonstrates how local program structures
interact with information flow directions.
In inference mode $\TI$, the type is always synthesised (i.e.,
determined by the term itself).
In checking mode $\TC$, the type is always inherited (i.e., determined
by the term context).
In function mode $\FN\M$, the type is a function type whose argument
type is bidirectional and whose result type depends on $\M$
recursively.

\begin{figure}[htb] \small
\raggedright
\boxed{\typp[\M]{\Gamma}{M}{\bid{A}}}
\hfill
\begin{mathparshrink}
\inferrule*[Lab=\dtylab{Unit}]
{~}
{\typi{\Gamma}{\eunit}{{\tunit}}}

\inferrule*[Lab=\dtylab{Var}]
{
  \Gamma\ni x : A \\
}
{\typi{\Gamma}{x}{\syn{A}}}

\inferrule*[Lab=\dtylab{Anno}]
{
  \typc{\Gamma}{M}{\inh{A}} \\
}
{\typi{\Gamma}{(M:A)}{\syn{A}}}

{
\inferrule*[Lab=\dtylab{Sub}]
{
  \typi{\Gamma}{M}{\syn{A}} \\
  \hl{\typsub[\M]{\Gamma}{\syn{A}}{\bid{B}}} \\
}
{\typp[\M]{\Gamma}{M}{\bid{B}}}
}

\inferrule*[Lab=\dtylab{Freeze}]
{
  \typi{\Gamma}{M}{\syn{A}} \\
}
{\typp[\TC]{\Gamma}{\Freeze{M}}{\inh{A}}}

\inferrule*[Lab=\dtylab{Forall}]
{
  \typc{\Gamma,\alpha}{V}{\inh{A}} \\ %
}
{\typc{\Gamma}{V}{\inh{\forall\alpha.A}}}

\inferrule*[Lab=\dtylab{Abs}]
{
  \typp[\M-]{\Gamma,x:A}{M}{\bid{B}} \\
}
{\typp[\M]{\Gamma}{\lambda x . M}{
  \inh{A} \to \bid{B}
}}

\inferrule*[Lab=\dtylab{AbsAnno}]
{
  \typp[\M-]{\Gamma,x:A}{M}{\bid{B}}
}
{\typp[\M]{\Gamma}{\lambda x^A . M}{
  \bid{A} \to \bid{B}
}}

{
\inferrule*[Lab=\dtylab{App}]
{
  \hl{\pati{\Gamma}{N}{\syn{P}}} \\
  \hl{\Gamma \vdash \inh{P} \fillsyn \bid{A}} \\\\
  \typp[\FN\M]{\Gamma}{M}{\bid{A}\to \bid{B}} \\
  \typc{\Gamma}{N}{\inh{A}} \\
}
{\typp[\M]{\Gamma}{M\;N}{\bid{B}}}
}
\end{mathparshrink}
\caption{Declarative typing rules for \Calc.}
\label{fig:decl-typing}
\end{figure}

Many of the rules are standard bidirectional typing rules.
Rule \dtylab{Unit} infers the unit type for unit.
Rule \dtylab{Var} infers the type $\syn{A}$ for variable $x$ from the
context.
Rule \dtylab{Anno} checks the term $M$ against the type $\inh{A}$
inherited from the type annotation and infers the type $\syn{A}$.
As mentioned in \Cref{sec:overview-colours}, the types $A$, $\inh{A}$,
and $\syn{A}$ in this rule are identical modulo colour.
Rule \dtylab{Forall} introduces an inherited universal quantifier in
checking mode.
We require the quantifier to be inherited to avoid guessing type
abstraction.

Upcasting and abstraction rules allow an arbitrary mode $\M$ in the
conclusion, generalising standard bidirectional typing, which only
allows checking mode.
Rule \dtylab{Sub} upcasts the synthesised type $\syn{A}$ of a term $M$
to the bidirectional type $\bid{B}$ guided by the mode $\M$.
We define the subtyping judgement in \Cref{sec:subtyping}.
Rule \dtylab{Abs} introduces an unannotated lambda abstraction.
We require the argument type to be inherited to avoid guessing
quantifiers without restricting it to monotypes.
Rule \dtylab{AbsAnno} introduces an annotated lambda abstraction.
The argument type in the conclusion does not need to be inherited
since it is fully determined by the known type annotation.
In both rules, we use $\M-$ as the mode of the premises, defined by
$(\FN\M)- = \M$ and as the identity on $\TI$ and $\TC$.

Since monotypes are uncoloured, our rule \dtylab{Abs} allows monotypes
to be guessed for arguments of lambda abstractions.
For instance, the type inference judgement
 ${\typp[\TI]{\cdot}{\lambda x . x}{\tunit \to \tunit}}$ can be derived by
 applying \dtylab{Abs} to
 ${\typp[\TI]{x:\tunit}{x}{\tunit}}$.

Rule \dtylab{App} is key to enabling bidirectional type information
flow, as we explained in detail in \Cref{sec:overview-application}.
We will formally define skeletons $P$, skeleton inference
$\pati{\Gamma}{N}{\syn{P}}$, and skeleton refinement $\Gamma\vdash
\inh{P}\fillsyn \bid{A}$ in
\Cref{sec:skeleton,sec:skeleton-inference}.

Rule \dtylab{Freeze} freezes the synthesised type $\syn{A}$ of term
$M$, as explained in \Cref{sec:overview-freezing}.
It disables further upcasting by switching from inference mode $\TI$
to checking mode $\TC$.
For instance, we cannot apply rule \dtylab{Sub} to the term
$\Freeze{\dec{id}}$, because $\Freeze{\dec{id}}$ must be in checking
mode, while $\dtylab{Sub}$ requires inference mode for its premise.
Similarly, we cannot apply rule \dtylab{Sub} to the term $(\lambda x .
\Freeze{id})$ either, which can only be in checking mode or function
modes.
Moreover, we may not apply \dtylab{Forall} to $\Freeze{\dec{id}}$, as
the syntax of values $V$ does not include frozen terms.

\subsection{Syntax-Directed Typing Rules}
\label{sec:syntax-directed-typing}

Our typing rules in \Cref{fig:decl-typing} enjoy a clear connection to
conventional bidirectional typing rules.
However, they are not syntax-directed because of overlaps introduced
by rules \dtylab{Sub} and \dtylab{Forall}.
We incorporate both rules into other rules to obtain a syntax-directed
type system as follows.
\[\small\ba{l@{\quad}l@{\quad}l@{\quad}l}
{
\inferrule*[Lab=\sdtylab{Unit}]
{
  \typsub[\M]{\Gamma}{\tunit}{\bid{A}} 
}
{\typp[\M]{\Gamma}{\eunit}{\bid{A}}}
}

&
{
\inferrule*[Lab=\sdtylab{Var}]
{
  \Gamma\ni x : A \quad
  \typsub[\M]{\Gamma}{\syn{A}}{\bid{B}}
}
{\typp[\M]{\Gamma}{x}{\bid{B}}}
}

&
\inferrule*[Lab=\sdtylab{Abs}]
{
  \M\neq\TC \quad
  \typp[\M-]{\Gamma,x:A}{M}{\bid{B}} \\
}
{\typp[\M]{\Gamma}{\lambda x . M}{
  \inh{A} \to \bid{B}
}}

&
\inferrule*[Lab=\sdtylab{AbsCheck}]
{
  \typc{\Gamma,\ol \alpha,x:A}{M}{\inh{B}} \\
}
{\typc{\Gamma}{\lambda x . M}{\inh{\forall\ol\alpha.}
  \inh{A} \to \inh{B}
}}
\\[2ex]

&
{
\inferrule*[Lab=\sdtylab{Anno}]
{
  \typp[\TC]{\Gamma}{M}{\inh{A}}\quad
  \typsub[\M]{\Gamma}{\syn{A}}{\bid{B}}
}
{\typp[\M]{\Gamma}{(M:A)}{\bid{B}}}
}

&
\inferrule*[Lab=\sdtylab{AbsAnno}]
{
  \M\neq\TC \quad
  \typp[\M-]{\Gamma,x:A}{M}{\bid{B}}
}
{\typp[\M]{\Gamma}{\lambda x^A . M}{
  \bid{A} \to \bid{B}
}}

&
\inferrule*[Lab=\sdtylab{AbsAnnoCheck}]
{
  \typc{\Gamma,\ol \alpha, x:A}{M}{\inh{B}}
}
{\typc{\Gamma}{\lambda x^A . M}{\inh{\forall\ol\alpha.}
  \inh{A} \to \inh{B}
}}
\ea\]
Concretely, we incorporate \dtylab{Sub} into the three base rules
\dtylab{Unit}, \dtylab{Var}, and \dtylab{Anno}, having the subtyping
judgement as a premise.
We incorporate \dtylab{Forall} into the two lambda abstraction rules
\dtylab{Abs} and \dtylab{AbsAnno}, which requires separating the rules
for checking mode \sdtylab{AbsCheck} and \sdtylab{AbsAnnoCheck}, where
we introduce inherited quantifiers, from other modes.

We prove in our Rocq mechanisation that the syntax-directed type
system is equivalent to the declarative type system with lemma
statements in \Cref{app:syntax-directed}.
This syntax-directed type system helps with the design of skeleton
inference in \Cref{sec:skeleton-inference} and the design of type
inference algorithms in \Cref{sec:algorithmic}.

\subsection{Subtyping and Consistency}
\label{sec:subtyping}

A fundamental principle of \Calc is to never guess polytypes out of
thin air.
Instead, we only guess polytypes that can be determined by inherited
type information.
In order to make sure that every universal quantifier we guess is
determined by some inherited quantifier, we decorate universal
quantifiers with tags as in the following syntax.
\[\small
\ba[t]{@{}=l@{\quad}+r@{~}+c@{~}+l@{}}
\slab{Decorated types} &\syn{\varphi} &::= &
    \tunit \mid {\alpha}
    \mid \syn{\varphi_1} \to \syn{\varphi_2}
    \mid \syn{\foralltag{\epsilon} \alpha.} \syn{\varphi}
    \\
\slab{Unresolved types} &\syn{T} &::= &
    \tunit \mid {\alpha}
    \mid \syn{T_1} \to \syn{T_2}
    \mid \syn{\foralltag{\delta} \alpha.} \syn{T}
    \\
\ea
\qquad
\ba[t]{@{}=l@{\quad}+r@{~}+c@{~}+l@{}}
\slab{Tag contexts} &\Psi &::=& \cdot \mid \Psi,\epsilon \\
\slab{Potential tags} &\delta &::=& \cdot \mid \epsilon \\
\ea
\qquad
\ba[t]{@{}=l@{\quad}+r@{~}+c@{~}+l@{}}
\slab{Tags} &\epsilon \\
\ea
\]

Tags track the provenance of quantifiers, as explained in
\Cref{sec:overview-subtyping}.
A decorated type $\syn{\varphi}$ is always synthesised and includes a
tag $\epsilon$ on every quantifier.
An unresolved type $\syn{T}$ is always synthesised and may have tags
on some quantifiers.
A tag context $\Psi$ is a set of tags.
We write $\Psi,\Psi'$ for set union and $\Psi\uplus\Psi'$ for
disjoint union.
A potential tag $\delta$ is either a tag or empty.
We define $\Psi,\delta$ as $\Psi$ if $\delta$ is empty, and
as $\Psi,\epsilon$ if $\delta$ is $\epsilon$.
We write $\tags{\Psi}{\syn{\varphi}}$ to mean that decorated type
$\syn{\varphi}$ is well-defined with respect to tag context $\Psi$,
that is, that $\Psi$ contains all tags in $\syn{\varphi}$ and each
quantifier has a unique tag:
\begin{mathparshrink} \small
  \inferrule*
  {~}
  {\tags{\cdot}{\syn{\tunit}}}

  \inferrule*
  {~}
  {\tags{\cdot}{\syn{\alpha}}}

  \inferrule*
  {\tags{\Psi}{\syn{\varphi_1}}  \\
    \tags{\Psi'}{\syn{\varphi_2}}}
  {\tags{\Psi\uplus \Psi'}{\syn{\varphi_1} \to \syn{\varphi_2}}}

  \inferrule*
  {\tags{\Psi}{\syn{\varphi}}}
  {\tags{\Psi \uplus \{\epsilon\}}{\syn{\foralltag{\epsilon}\alpha.\varphi}}}
\end{mathparshrink}
We write $\Gamma\vdash\tags{\Psi}{\varphi}$ as shorthand for
$\Gamma\vdash\syn{\varphi}$ and $\tags{\Psi}{\syn{\varphi}}$, where
$\Gamma\vdash\syn{\varphi}$ is standard well-scopedness.

\Cref{fig:decl-subtyping-consistency} defines the subtyping and
consistency relation for \Calc.
The subtyping judgement has the form
$\typsub[\M][\Psi]{\Gamma}{\syn{T}}{\bid{A}}$ (abbreviated as
$\typsub[\M]{\Gamma}{\syn{T}}{\bid{A}}$ if $\Psi$ is empty), which
states that given tags $\Psi$ under context $\Gamma$ in mode $\M$,
unresolved type $\syn{T}$ is a subtype of bidirectional type
$\bid{A}$.
As with the typing judgement, we require standard well-scopedness and
$\skmod[\TY]{\bid{A}}{\M}$ as well-formedness conditions for the
subtyping judgement.
The consistency judgement has the form
$\typconsis[\Psi]{\Gamma}{\syn{T}}{\bid{A}}$, which states that given
tags $\Psi$ under context $\Gamma$ the unresolved type $\syn{T}$ is
consistent with the bidirectional type $\bid{A}$.
Intuitively, in addition to their usual roles, subtyping and
consistency rules aim to use inherited quantifiers from $\bid{A}$ to
determine tagged quantifiers in $\syn{T}$, so that these tagged
quantifiers are not guessed out of thin air.
The $\Psi$ does not contain all tags in $\syn{T}$, but only those
determined from $\bid{A}$.

{
\makeatletter
\def\TirNameStyle#1{\footnotesize\scshape #1}
\def\LeftTirNameStyle#1{\TirNameStyle{#1}}
\def\RightTirNameStyle#1{\TirNameStyle{#1}}
\makeatother
\begin{figure}[htbp]\footnotesize
\raggedright
\boxed{\typsub[\M][\Psi]{\Gamma}{\syn{T}}{\bid{A}}}
\hfill
\begin{mathparshrink}
\inferrule*[Lab=\sublab{Infer}]
{
  \typconsis[\cdot]{\Gamma}{\syn{T}}{\syn{A}}
}
{\typsub[\TI][\cdot]{\Gamma}{\syn{T}}{\syn{A}}}
\quad\!
\inferrule*[Lab=\sublab{Check}]
{
  \syn{T},\inh{A}\text{ guarded} \\\\
  \typconsis[\Psi]{\Gamma}{\syn{T}}{\inh{A}}
}
{\typsub[\TC][\Psi]{\Gamma}{\syn{T}}{\inh{A}}}
\quad\!
\inferrule*[Lab=\sublab{ForallR}]
{
  \typsub[\TC][\Psi]{\Gamma,\alpha}{\syn{T}}{\inh{A}}
}
{\typsub[\TC][\Psi]{\Gamma}{\syn{T}}{\inh{\forall\alpha.}\inh{A}}}
\quad\!
\inferrule*[Lab=\sublab{Arrow}]
{
  \typconsis[\Psi_1]{\Gamma}{\syn{T_1}}{\bid{A_1}} \\\\
  \typsub[\M][\Psi_2]{\Gamma}{\syn{T_2}}{\bid{A_2}}
}
{\typsub[\FN\!\M][\Psi_1,\!\Psi_2]{\Gamma}{
  \syn{T_1}\!\!\to\!\! \syn{T_2}\!}{\!\bid{A_1} \!\!\to\!\! \bid{A_2}}}
\quad\!
\inferrule*[Lab=\sublab{ForallL}]
{
  \M\neq\TI \quad %
  \Psi' \text{ fresh} \quad
  \Gamma \vdash \tags{\Psi'}{\syn{\varphi}} \\\\
  \typsub[\M][\Psi,\Psi']{\Gamma}{\syn{T}[\syn{\varphi}/\alpha]}{\bid{A}} \\
}
{\typsub[\M][\Psi]{\Gamma}{\syn{\foralltag{\delta} \alpha. T}}{\bid{A}}}
\end{mathparshrink}

\raggedright
\boxed{\typconsis[\Psi]{\Gamma}{\syn{T}}{\bid{A}}}
\hfill
\vspace{-.5em}
\begin{mathparshrink}
\inferrule*[Lab=\eqlab{Unit}]
{~}
{\typconsis[\cdot]{\Gamma}{\tunit}{\tunit}}
\quad
\inferrule*[Lab=\eqlab{Var}]
{
  \Gamma\ni\alpha
}
{\typconsis[\cdot]{\Gamma}{\alpha}{\alpha}}
\quad
\inferrule*[Lab=\eqlab{ForallInh}]
{
  \typconsis[\Psi]{\Gamma,\alpha}{\syn{T}}{\bid{A}} \\
}
{\typconsis[\Psi,\delta]{\Gamma}
  {\syn{\foralltag{\delta}\alpha.T}}{\inh{\forall\alpha.} \bid{A}}
  }
\quad
\inferrule*[Lab=\eqlab{ForallSyn}]
{
  \typconsis[\Psi]{\Gamma,\alpha}{\syn{T}}{\bid{A}} \\
}
{\typconsis[\Psi]{\Gamma}
  {\syn{\foralltag{\delta}\alpha.T}}{\syn{\forall\alpha.} \bid{A}}
  }
\quad
\inferrule*[Lab=\eqlab{Arrow}]
{
  \typconsis[\Psi_1]{\Gamma}{\syn{T_1}}{\bid{A_1}} \\\\
  \typconsis[\Psi_2]{\Gamma}{\syn{T_2}}{\bid{A_2}}
}
{\typconsis[\Psi_1, \Psi_2]{\Gamma}{\syn{T_1} \to \syn{T_2}}{\bid{A_1} \to \bid{A_2}}}
\end{mathparshrink}

\caption{Subtyping and consistency rules for \Calc.}
\label{fig:decl-subtyping-consistency}
\end{figure}
}

Rule \sublab{Infer} solves no tags in inference mode.
Rule \sublab{Check} degenerates to consistency when both sides are
guarded. %
Rule \sublab{ForallR} introduces an inherited quantifier in checking
mode.
Rule \sublab{Arrow} upcasts a function type $\syn{T_1}\to\syn{T_2}$ to
a bidirectional function type $\bid{A_1} \to\bid{A_2}$ in function
mode $\FN\M$ by unifying the argument types and recursively upcasting
the result type.
This does not break our claim of only allowing shallow subtyping,
because the mode $\FN\M$ indicates that the term of this function type
is indeed applied to an argument, so the upcasting of its result type
is allowed.
The solved tags of the conclusion are the union of those from the two
premises.
In this sense, the tag context is a relevant context: tags may be
duplicated, but each tag it contains must be determined.

Rule \sublab{ForallL} is the most interesting one, since it is where
guessing polytypes happens.
We have explained its core idea with the motivating example in
\Cref{sec:overview-subtyping}.
It requires the mode not to be $\TI$ to avoid overlapping with
\sublab{Infer} where no instantiation is allowed.
For instance, the type $\syn{\forall\alpha}.\alpha\to\alpha$ can only
be upcast to itself in inference mode; we cannot instantiate it.
For the instantiation, we guess a decorated type $\syn{\varphi}$ whose
tags in $\Psi'$ are unique and fresh.
Adding all tags $\Psi'$ of $\syn{\varphi}$ to the tag context
of the premise subtyping judgement ensures that every quantifier in
$\syn{\varphi}$ is determined by inherited information from $\bid{A}$.
The potential tag $\delta$ is not solved in this rule since we cannot
guarantee that the quantifier being instantiated must exist.

Most consistency rules are straightforward.
Rule \eqlab{ForallInh} is the only rule where we add a tag to the tag
context, since this tag is determined by the inherited quantifier on
the RHS.
Rule \eqlab{ForallSyn} does not determine any tag since the RHS is
synthesised.

\subsection{Skeletons and Ghosts}
\label{sec:skeleton}

Skeletons extend types with ghosts to represent unknown type information.
\[\small\ba{@{}=l@{\mkern10mu}+r@{~}+c@{~}+l@{}}
  \slab{Skeletons} &P,Q &::= &
    \tunit \mid \alpha \mid P\to Q
    \mid \forall\alpha.P
    \phantom{\mid \inh{\forall\alpha .} \inh{P}}\;
    \mid {\Hole}
    \mid {\TopHole. P}
    \\
  \slab{Synthesised skeletons}  &\syn{P}, \syn{Q} &::= &
    \tunit \mid {\alpha} \mid \syn{P} \to \syn{Q}
    \mid \syn{\forall\alpha .} \syn{P}
    \phantom{\mid \inh{\forall\alpha .} \inh{P}}\;
    \mid {\Hole}
    \mid \TopHole . \syn{P}
    \\
  \slab{Inherited skeletons}  &\inh{P}, \inh{Q} &::= &
    \tunit \mid {\alpha} \mid \inh{P} \to \inh{Q}
    \mid \inh{\forall\alpha .} \inh{P}
    \phantom{\mid \inh{\forall\alpha .} \inh{P}}\;
    \mid {\Hole}
    \mid \TopHole . \inh{P}
    \\
  \slab{Bidirectional skeletons}  &\bid{P}, \bid{Q} &::= &
    \tunit \mid {\alpha} \mid \bid{P} \to \bid{Q}
    \mid \syn{\forall\alpha .} \bid{P}
    \mid \inh{\forall\alpha .} \bid{P}
    \mid {\Hole}
    \mid {\TopHole .} \bid{P}
    \\
\ea\]

A plain ghost $\Hole$ represents an unknown type.
A universal ghost $\TopHole$ in $\TopHole. P$ represents a sequence of
unknown universal quantifiers.
Like types, skeletons may be coloured as synthesised skeletons
($\syn{P}$), inherited skeletons ($\inh{P}$), and bidirectional
skeletons ($\bid{P}$).
Only concrete quantifiers are coloured; ghosts themselves are always
black. The guarded condition is extended to skeletons, with $\TopHole.
P$ and $\Hole$ being unguarded, and similarly extended to coloured
skeletons.

\paragraph{Well-formedness of skeletons}

We require universal ghosts to only appear at the end of a sequence of
quantifiers.
This is important for our algorithm to deterministically solve the
unification between two quantifier sequences.
Moreover, given a skeleton $P$, for each universal ghost $\TopHole.Q$
in $P$, a type variable free in $Q$ cannot be bound in $P$.
This maintains the above invariant under instantiation.
For instance, under a context containing $\beta$,
$\forall\alpha.\TopHole.\alpha$ is not well-formed, because the bound
variable $\alpha$ appears under a universal ghost, while
$\forall\alpha.\TopHole.\beta$ is well-formed, because $\beta$ is not
bound by the skeleton but by the context.

We characterise these requirements on universal ghosts via the
well-formedness of skeletons.
We write $\Gamma \vdash \bid{P}$ if the bidirectional skeleton
$\bid{P}$ is well-formed under context $\Gamma$.
We use an auxiliary set $\Delta$ to track the type variables bound
by universal quantifiers in the skeleton, since these variables cannot
be used under universal ghosts.
We write $\Gamma\setminus\Delta$ for the context after removing the
variables in $\Delta$ from $\Gamma$.
We define the judgement, whose full form is $\Gamma\mid\Delta \vdash
\bid{P}$, as follows.
{
  \newcommand{\margin}{\kern 1.2em}
  \begin{mathparshrink} \small
    \inferrule*
    { }
    {\Gamma\mid\Delta \vdash \Hole}
    \margin
    \inferrule*
    {~}
    {\Gamma\mid\Delta \vdash \TUnit}
    \margin
    \inferrule*
    {\Gamma \ni \alpha}
    {\Gamma\mid\Delta \vdash \alpha}
    \margin
    \inferrule*
    { \Gamma\mid\Delta \vdash \bid{P} \margin
      \Gamma\mid\Delta \vdash \bid{Q}}
    {\Gamma\mid\Delta \vdash \bid{P}\to \bid{Q}}
    \margin
    \inferrule*
    {\Gamma,\alpha\mid\Delta,\alpha \vdash \bid{P}}
    {\Gamma\mid\Delta \vdash \bid{\forall\alpha.P}}
    \margin
    \inferrule*
    {
      \guarded{\bid{P}} \margin
      \Gamma\setminus\Delta\mid\cdot \vdash \bid{P}
    }
    {\Gamma\mid\Delta \vdash \TopHole. \bid{P}}
  \end{mathparshrink}
}

We write $\bid{\forall\alpha}$ to match both $\syn{\forall\alpha}$ and
$\inh{\forall\alpha}$.
Since colours play no role in the above rules, we define
well-formedness of uncoloured skeletons ($\Gamma\vdash P$) by simply
ignoring colours in these rules.

\paragraph{Skeleton refinement}
We define a skeleton refinement relation
$\Gamma\vdash\inh{P}\fillsyn\bid{Q}$ which fills in ghosts on the LHS
with synthesised skeletons to obtain the RHS.
We require well-formedness of both $\inh{P}$ and $\bid{Q}$ under
$\Gamma$ as well-formedness conditions of the refinement judgement.
Skeleton refinement characterises what skeletons a skeleton can
represent.
As discussed in \Cref{sec:overview-colours}, given a skeleton $P$, for
every ghost in $P$, it cannot represent skeletons that use type
variables bound in $P$.
To enforce this requirement, we augment the full judgement form to
$\skctx[\Delta]{\Gamma} \inh{P}\fillsyn \bid{Q}$, where the set
$\Delta$ tracks type variables bound in $P$.
We define the judgement as follows.
\begin{mathparshrink} \rulesize
\inferrule*
{~}
{\skctx[\Delta]{\Gamma} \tunit \fillsyn \tunit}

\inferrule*
{
  ~
}
{\skctx[\Delta]{\Gamma} \alpha \fillsyn \alpha}

\inferrule*
{
  \skctx[\Delta]{\Gamma} \inh{P_1} \fillsyn \bid{Q_1} \\
  \skctx[\Delta]{\Gamma} \inh{P_2} \fillsyn \bid{Q_2}
}
{\skctx[\Delta]{\Gamma}
  \inh{P_1}\to \inh{P_2} \fillsyn \bid{Q_1}\to \bid{Q_2}}

\inferrule*
{
  \skctx[\Delta,\alpha]{\Gamma,\alpha} \inh{P}\fillsyn \bid{Q}
}
{\skctx[\Delta]{\Gamma}
  \inh{\forall\alpha.P} \fillsyn \inh{\forall\alpha.}\bid{Q}}

\inferrule*
{
  \Delta\cap\ftv{\syn{Q}} = \emptyset
  }
{\skctx[\Delta]{\Gamma} \Hole \fillsyn \syn{Q}}

\inferrule*
{
  \skctx[\Delta]{\Gamma} \inh{P} \fillsyn \bid{Q}
}
{\skctx[\Delta]{\Gamma} \TopHole. \inh{P} \fillsyn \bid{Q}}

\inferrule*
{
  \skctx[\Delta]{\Gamma,\alpha} %
    \TopHole.\inh{P} \fillsyn \bid{Q}
}
{\skctx[\Delta]{\Gamma}
  \TopHole. \inh{P} \fillsyn \syn{\forall\alpha.} \bid{Q}}

\inferrule*
{
  \skctx[\Delta]{\Gamma} \inh{P} \fillsyn \bid{Q}
}
{\skctx[\Delta]{\Gamma}
  \TopHole. \inh{P} \fillsyn \TopHole. \bid{Q}}
\end{mathparshrink}

As with well-formedness, we define the refinement of uncoloured
skeletons ($\Gamma\vdash P\fill{} Q$) by ignoring colours in the above
rules.
For refining $\Hole$, we require $\syn{Q}$ to contain no variables
from $\Delta$.
For refining $\TopHole$, we have three rules.
The first rule removes a $\TopHole$ on the left.
The second rule feeds a concrete quantifier $\alpha$ on the right into
a $\TopHole$ on the left.
We do not add $\alpha$ to $\Delta$ in the premise because it is bound
by the synthesised quantifier and can be used when filling ghosts,
unlike variables bound by inherited quantifiers in $\Delta$.
The third rule removes a $\TopHole$ on the left and the right.

\subsection{Skeleton Inference}
\label{sec:skeleton-inference}
There are different skeleton inference strategies.
The most naive approach is to always infer the trivial skeleton
$\Hole$, which provides no useful information.
A more refined and principled approach to skeleton inference is to
follow the structure of the typing rules, inserting ghosts as needed.

To define skeleton inference, we extend the syntax of variable
bindings $x:A$ in contexts $\Gamma$ to $x:P$.
\Cref{fig:decl-skeleton-inference} gives the skeleton inference rules.
Skeleton inference has the form $\patp[\M]{\Gamma}{M}{\bid{P}}$, which
means that the term $M$ has skeleton $\bid{P}$ under context $\Gamma$
in mode $\M$.
We require well-formedness of $M$ and $\bid{P}$ under $\Gamma$ as
well-formedness conditions for the skeleton inference judgement, as
well as the compatibility between bidirectional skeletons and modes
$\skmod[\SK]{\bid{P}}{\M}$.
The relation $\skmod[\SK]{\bid{P}}{\M}$ is defined similarly to
$\skmod[\TY]{\bid{A}}{\M}$. The only difference is for the checking
mode: $\skmod[\SK]{\syn{P}}{\TC}$.

\begin{figure}[htbp] \small
\raggedright
\boxed{\patp[\M]{\Gamma}{M}{\bid{P}}}
\hfill
\begin{mathparshrink}
\inferrule*[Lab=\plab{Unit}]
{\patsub[\M]{\Gamma}{\tunit}{\bid{P}}}
{\patp[\M]{\Gamma}{\eunit}{\bid{P}}}

\inferrule*[Lab=\plab{Var}]
{
  \Gamma\ni x : P \\
  \patsub[\M]{\Gamma}{\syn{P}}{\bid{Q}}
}
{\patp[\M]{\Gamma}{x}{\bid{Q}}}

\inferrule*[Lab=\plab{Anno}]
{
  \patsub[\M]{\Gamma}{\syn{A}}{\bid{P}}
}
{\patp[\M]{\Gamma}{(M:A)}{\bid{P}}}

\inferrule*[Lab=\plab{Freeze}]
{
  \ngti{\Gamma}{M}{\syn{P}} \\
}
{\patp[\TC]{\Gamma}{\Freeze{M}}{\syn{P}}}

\inferrule*[Lab=\plab{Abs}]
{
  \M \neq \TC \\
  \patp[\M-]{\Gamma,x:P}{M}{\bid{Q}} \\
}
{\patp[\M]{\Gamma}{\lambda x . M}{
  \inh{P} \to \bid{Q}
}}

\inferrule*[Lab=\plab{AbsAnno}]
{
  \M \neq \TC \\
  \patp[\M-]{\Gamma,x:A}{M}{\bid{Q}}
}
{\patp[\M]{\Gamma}{\lambda x^A . M}{
  \bid{A} \to \bid{Q}
}}

\inferrule*[Lab=\plab{AbsCheck}]
{
  \pati{\Gamma,x:\Hole}{M}{\syn{Q}} \\
}
{\pati{\Gamma}{\lambda x . M}{
  {\TopHole . \Hole} \to \syn{Q}
}}

\inferrule*[Lab=\plab{AbsAnnoCheck}]
{
  \pati{\Gamma,x:A}{M}{\syn{Q}}
}
{\pati{\Gamma}{\lambda x^A . M}{
  {\TopHole .} \syn{A} \to \syn{Q}
}}

{
\inferrule*[Lab=\plab{App}]
{
  {\pati{\Gamma}{N}{\syn{P}}} \\
  \Gamma \vdash \inh{P} \fillsyn \bid{P'} \\
  \patp[\FN\M]{\Gamma}{M}{\bid{P'}\to \bid{Q}} \\
}
{\patp[\M]{\Gamma}{M\;N}{\bid{Q}}}
}
\end{mathparshrink}
\caption{Skeleton inference rules for \Calc.}
\label{fig:decl-skeleton-inference}
\end{figure}

Skeleton inference rules correspond one-to-one with the syntax-directed
variant of the typing rules in
\Cref{sec:syntax-directed-typing,sec:typing}.
There are two main differences.
The first is that in rules \plab{Anno} and \plab{App} we do not need
to do type checking, because skeleton inference only cares about
inferring a skeleton for the term, rather than ensuring that the term
is well-typed.
The second is that in \plab{AbsCheck} and \plab{AbsAnnoCheck} we
insert universal ghosts, while in the corresponding syntax-directed
typing rules \sdtylab{AbsCheck} and \sdtylab{AbsAnnoCheck} we have
inherited universal quantifiers.
We separate typing and skeleton inference rules only for
presentational purposes.

In \plab{Unit}, \plab{Var}, and \plab{Anno}, we use the subtyping
judgement between skeletons $\patsub[\M]{\Gamma}{\syn{P}}{\bid{Q}}$
which also mimics the subtyping judgement for types in
\Cref{sec:subtyping}.
Unlike typing, a general subsumption rule like \dtylab{Sub} does not
work for skeleton inference.
Otherwise, for example, we can infer a skeleton $\TopHole.\tunit \to (\forall
\alpha.\alpha \to \alpha)$ for $\lambda x^\tunit. \dec{id}$ under
mode $\TC$, by first inferring the skeleton of this term under mode
$\TI$ and then applying a general subsumption rule.
However, this skeleton is not compatible with instantiation of
$\dec{id}$.
The expected skeleton is $\TopHole.\tunit \to (\TopHole. \Hole \to \Hole)$.

\subsection{Skeleton Subtyping and Consistency}
\label{sec:skeleton-subtyping-consistency}
As in \Cref{sec:subtyping}, we define decorated skeletons and unresolved
skeletons as follows.
\[\small\ba{@{}=l@{\quad}+r@{~}+c@{~}+l@{}}
\slab{Decorated skeletons} &\syn{\psi} &::= &
    \tunit \mid {\alpha}
    \mid \syn{\psi_1} \to \syn{\psi_2}
    \mid \syn{\foralltag{\epsilon} \alpha.} \syn{\psi}
    \mid {\HoleG}
    \mid {\TopHoleG} . \syn{\psi}
    \\
\slab{Unresolved skeletons} & \syn{S} &::= &
    \tunit \mid {\alpha}
    \mid \syn{S_1} \to \syn{S_2}
    \mkern 3mu\mid \syn{\foralltag{\delta} \alpha.}\syn{S}
    \mid {\Hole}
    \mid {\HoleG}
    \mid {{\TopHole . \syn{S}}
    \mid {\TopHoleG} . \syn{S}}
    \\
\ea\]
In addition to adding tags to quantifiers, we also use dotted ghosts
$\HoleG$ and $\TopHoleG$ in decorated skeletons.
Dotted ghosts are used to avoid trivial guesses for instantiation as
explained later.
As in \Cref{sec:subtyping}, we write $\tags{\Psi}{\syn{\psi}}$ if
$\Psi$ contains all tags in $\syn{\psi}$ and each quantifier has a
unique tag.
We define an erasure operation $\erase{S}$ which gives a synthesised
skeleton by removing all tags and transforming dotted ghosts to normal
ghosts.
Well-formedness and guardedness of decorated skeletons and unresolved
skeletons are defined by those of their erasure.

\Cref{fig:decl-skeleton-subtyping-consistency} defines
skeleton subtyping and skeleton consistency.
Most of the subtyping and consistency rules mimic their type
counterparts in \Cref{fig:decl-subtyping-consistency}.
We highlight the new rules for ghosts.
In \Cref{app:all-in-one-subtyping}, we combine subtyping for types and
skeletons into one judgement
$\typsub[\N][\M][]{\Gamma}{\syn{P}}{\bid{Q}}$ with the sort $\N = \TY
\mid \SK$ indicating whether the judgement is for types or skeletons.

\begin{figure}[htbp]\footnotesize
\makeatletter
\def\TirNameStyle#1{\footnotesize\scshape #1}
\def\LeftTirNameStyle#1{\TirNameStyle{#1}}
\def\RightTirNameStyle#1{\TirNameStyle{#1}}
\makeatother

\raggedright
\boxed{\patsub[\M][\Psi]{\Gamma}{\syn{S}}{\bid{P}}}
\hfill
\begin{mathparshrink}
\inferrule*[Lab=\sublab{Infer}]
{
  \typconsis[\cdot]{\Gamma}{\syn{S}}{\syn{P}}
}
{\patsub[\TI][\cdot]{\Gamma}{\syn{S}}{\syn{P}}}

\hl{
\inferrule*[Lab=\sublab{UnivGhostR}]
{
  \syn{S},\syn{P}\text{ guarded} \quad
  \typconsis[\cdot]{\Gamma}{\syn{S}}{\syn{P}}
}
{\patsub[\TC][\cdot]{\Gamma}{\syn{S}}{\TopHole . \syn{P}}}
}

\hl{
\inferrule*[Lab=\sublab{Ghost}]
{
  ~
}
{\patsub[\M][\cdot]{\Gamma}{{\Hole}}{\inh{P}}}
}

\hl{
\inferrule*[Lab=\sublab{GhostG}]
{
  ~
}
{\patsub[\M][\cdot]{\Gamma}{{\HoleG}}{\Hole}}
}

\inferrule*[Lab=\sublab{Arrow}]
{
  \typconsis[\Psi_1]{\Gamma}{\syn{S_1}}{\bid{P_1}} \\\\
  \patsub[\M][\Psi_2]{\Gamma}{\syn{S_2}}{\bid{P_2}}
}
{\patsub[\FN\M][\Psi_1,\Psi_2]{\Gamma}{\syn{S_1}\to \syn{S_2}}{\bid{P_1} \to \bid{P_2}}}
\quad\
\inferrule*[Lab=\sublab{ForallL}]
{
  \M\neq\TI \\
  \Psi' \text{ fresh} \\
  {\Gamma} \vdash \tags{\Psi'}{\syn{\psi}} \\\\
  \patsub[\M][\Psi,\Psi']{\Gamma}{\syn{S}[\syn{\psi}/\alpha]}{\bid{P}} \\
}
{\patsub[\M][\Psi]{\Gamma}{\syn{\foralltag{\delta} \alpha. S}}{\bid{P}}}
\quad\
\hl{
\inferrule*[Lab=\sublab{UnivGhostL}]
{
  \M\neq\TI \\\\
  \patsub[\M][\Psi]{\Gamma}{\syn{S}}{\bid{P}}
}
{\patsub[\M][\Psi]{\Gamma}{\TopHole. \syn{S}}{\bid{P}}}
\quad\
\inferrule*[Lab=\sublab{UnivGhostLG}]
{
  \M\neq\TI \\\\
  \patsub[\M][\Psi]{\Gamma}{\syn{S}}{\bid{P}}
}
{\patsub[\M][\Psi]{\Gamma}{\TopHoleG. \syn{S}}{\bid{P}}}
}
\end{mathparshrink}

\raggedright
\boxed{\typconsis[\Psi]{\Gamma}{\syn{S}}{\bid{P}}}
\hfill
\vspace{-.7\baselineskip}
\begin{mathparshrink}
\inferrule*[Lab=\eqlab{Unit}]
{~}
{\typconsis[\cdot]{\Gamma}{\tunit}{\tunit}}
\quad\
\inferrule*[Lab=\eqlab{Var}]
{
  \Gamma\ni\alpha
}
{\typconsis[\cdot]{\Gamma}{\alpha}{\alpha}}
\quad\
\inferrule*[Lab=\eqlab{ForallInh}]
{
  \typconsis[\Psi]{\Gamma,\alpha}{\syn{S}}{\bid{P}} \\
}
{\typconsis[\Psi,\delta]{\Gamma}
  {\syn{\foralltag{\delta}\alpha.S}}{\inh{\forall\alpha.} \bid{P}}
  }
\quad\
\inferrule*[Lab=\eqlab{ForallSyn}]
{
  \typconsis[\Psi]{\Gamma,\alpha}{\syn{S}}{\bid{P}} \\
}
{\typconsis[\Psi]{\Gamma}
  {\syn{\foralltag{\delta}\alpha.S}}{\syn{\forall\alpha.} \bid{P}}
  }
\quad\
\inferrule*[Lab=\eqlab{Arrow}]
{
  \typconsis[\Psi_1]{\Gamma}{\syn{S_1}}{\bid{P_1}} \\\\
  \typconsis[\Psi_2]{\Gamma}{\syn{S_2}}{\bid{P_2}}
}
{\typconsis[\Psi_1, \Psi_2]{\Gamma}{\syn{S_1 \to S_2}}{\bid{P_1 \to P_2}}}

\hl{
\inferrule*[Lab=\eqlab{Ghost}]
{~}
{\typconsis[\cdot]{\Gamma}{{\Hole}}{\inh{P}}}
\quad
\inferrule*[Lab=\eqlab{GhostG}]
{
  ~
}
{\typconsis[\cdot]{\Gamma}{{\HoleG}}{\Hole}}
\quad
\inferrule*[Lab=\eqlab{UnivGhostG}]
{
  \typconsis[\Psi]{\Gamma}{\syn{S}}{\bid{P}}
}
{\typconsis[\Psi]{\Gamma}{\TopHoleG . \syn{S}}{{\TopHole.}{\bid{P}}}}
\quad
\inferrule*[Lab=\eqlab{UnivGhost}]
{
  \typconsis[\Psi]{\Gamma}{\syn{S}}{\bid{P}}
}
{\typconsis[\Psi]{\Gamma}{\TopHole. \syn{S}}{{\TopHole.}{\bid{P}}}}
\quad
\inferrule*[Lab=\eqlab{UnivGhost1}]
{
  \typconsis[\Psi]{\Gamma,\alpha}{\TopHole.\syn{S}}{\bid{P}}
}
{\typconsis[\Psi]{\Gamma}{\TopHole. \syn{S}}{\inh{\forall\alpha.}{\bid{P}}}}
\quad
\inferrule*[Lab=\eqlab{UnivGhost2}]
{
  \typconsis[\Psi]{\Gamma}{\syn{S}}{\bid{P}}
}
{\typconsis[\Psi]{\Gamma}{\TopHole. \syn{S}}{{\bid{P}}}}
}
\end{mathparshrink}
\caption{Subtyping and consistency rules of skeletons for \Calc.}
\label{fig:decl-skeleton-subtyping-consistency}
\vspace{-.4\baselineskip}
\end{figure}

Skeleton subtyping $\patsub[\M][\Psi]{\Gamma}{\syn{S}}{\bid{P}}$ requires
well-formedness of $\syn{S}$ and $\bid{P}$ under $\Gamma$ as well as
$\skmod[\SK]{\bid{P}}{\M}$ as well-formedness conditions.
Intuitively, skeleton subtyping upcasts $\syn{S}$ to $\bid{P}$ guided
by inherited information in $\bid{P}$.
Rule \sublab{UnivGhostR} inserts a universal ghost in checking mode,
replacing \sublab{Check} and \sublab{ForallR} in
\Cref{fig:decl-subtyping-consistency}.
Rule \sublab{UnivGhostL} instantiates a universal ghost.
Like \sublab{ForallL}, it requires the mode not to be inference mode.
Rule \sublab{Ghost} upcasts a ghost to any inherited skeleton.
There are two new rules for dotted ghosts.
\sublab{UnivGhostLG} is the same as \sublab{UnivGhostL}.
\sublab{GhostG} forces the RHS to be a ghost to avoid trivial guesses.

Skeleton consistency $\typconsis[\Psi]{\Gamma}{\syn{S}}{\bid{P}}$
requires well-formedness of $\syn{S}$ and $\bid{P}$ under $\Gamma$ as
well as the refinement judgement $\Gamma\vdash \erase{S}\fill P$.
Intuitively, skeleton consistency refines $\syn{S}$ to $\bid{P}$
guided by inherited information in $\bid{P}$.
The first five rules are the same as their type counterparts in
\Cref{fig:decl-subtyping-consistency}, except that they are lifted to
skeletons.
The remaining rules are for ghosts and dotted ghosts.
Rule \eqlab{Ghost} matches a ghost with any inherited skeleton, while
rule \eqlab{GhostG} matches a dotted ghost with a ghost.
Rule \eqlab{UnivGhostG} matches a dotted universal ghost with a
universal ghost.
Rule \eqlab{UnivGhost}, \eqlab{UnivGhost1}, and \eqlab{UnivGhost2}
match a universal ghost to a universal ghost or an inherited quantifier,
or remove a universal ghost.

As shown in rules \sublab{GhostG}, \eqlab{GhostG}, and
\eqlab{UnivGhostG}, a dotted ghost can only match a ghost.
This avoids trivial guesses in \sublab{ForallL}.
Otherwise, for example, we can always guess the trivial skeleton
$\Hole$ for instantiation, which always succeeds but provides no
information.

\FloatBarrier
\section{Algorithmic Type System}
\label{sec:algorithmic}

The declarative type system in \Cref{sec:declarative} is
non-algorithmic because of guesses: we guess monotypes arbitrarily and
we guess decorated polytypes for instantiation in subtyping.
In this section, we give an algorithmic type system for \Calc
presented in the style of the type-inference-in-context
approach~\citep{Gundry13,GundryMM10}.
For guessing monotypes, we introduce flexible type variables and solve
them with standard first-order unification.
For guessing decorated polytypes, we introduce a novel look judgement
which first collects local polymorphic constraints for a type variable
and then constructs a polytype to instantiate the type variable.
This judgement is the most interesting machinery of this section, and
we present it as part of algorithmic subtyping in
\Cref{sec:algo-subtyping}.
We prove the algorithmic type system is sound and complete with
respect to the declarative type system in \Cref{sec:declarative}.

\subsection{Algorithmic Syntax and Information Increase}
\label{sec:alg-syntax}

We extend the syntax of types and skeletons with \emph{flexible} type
variables $\hat\alpha$ which stand for unsolved monomorphic types.
These variables are generated and manipulated by the algorithm and can
be unified with other monomorphic types.
In contrast, \emph{rigid} type variables come from the object language
and cannot be unified with anything.
Restricting them to monomorphic types is crucial to avoid guessing
polymorphic types.
The extended syntax for types, skeletons, and algorithmic contexts is
given below.
\[\small
\ba[t]{@{}=l@{\mkern9mu}+r@{~}+c@{~}+l@{}}
  \slab{Types}    &A,B &::=& \tunit\mid\alpha \mid A \to B \mkern8mu\mid \forall\alpha . A \mid \hl{\hat{\alpha}} \\
  \slab{Monotypes}&\tau,\pi &::=& \tunit\mid\alpha \mid \tau\to\pi \mkern11mu
  \phantom{\mid\forall\alpha.A}\mkern5mu
  \mid \hl{\hat{\alpha}} \\
  \slab{Skeletons} &P,Q &::=& \tunit\mid\alpha \mid P \to Q \mkern8mu\mid \forall\alpha.P
    \mid \hl{\hat{\alpha}}
    \mid \Hole \mid \TopHole. P
    \\
\ea
\ \ \
\ba[t]{@{}=l@{\mkern9mu}+r@{~}+c@{~}+l@{}}
  \slab{Entries} &e &::= &
    \alpha \mid x : P
    \mid \hl{\hat\alpha}
    \mid \hl{\hat\alpha \assign \tau}
    \\
  \slab{Algorithmic contexts}\! &\Theta &::=&
    \cdot \mid \Theta, e \vphantom{\hl{\hat{\alpha}}}\\
  \slab{Suffixes} &\Xi &::=& \cdot
    \mid {\Xi, \hat{\alpha}}
    \mid {\Xi, \hat{\alpha}\assign \tau} \vphantom{\hl{\hat{\alpha}}}
    \\
\ea
\]

Algorithmic contexts $\Theta$ are ordered. Entries can only depend on
those before them.
Flexible variables in algorithmic contexts $\Theta$ are either
declarations ($\hat\alpha$) or definitions ($\hat\alpha \assign
\tau$).
Suffixes $\Xi$ are lists of declarations and definitions of flexible
variables that can be moved around by algorithms.

Following \citet{GundryMM10}, we define metasubstitutions (or
information increase) as follows.
\begin{mathparshrink}\footnotesize
\inferrule*
{ }
{\inc[\iota]{\cdot}{\Xi}}
\quad
\inferrule*
{
  \inc[\theta]{\Theta}{\Theta'} \\
  \Theta' \vdash \tau
}
{\inc[(\theta,\tau/\hat\alpha)]{\Theta,{\hat\alpha}}{\Theta'}}
\quad
\inferrule*
{
  \inc[\theta]{\Theta}{\Theta'} \\
  \Theta'\vdash \tyequiv{\tau'}{\theta \tau}
}
{\inc[(\theta,\tau'/\hat\alpha)]{\Theta,{\hat\alpha}\assign{\tau}}{\Theta'}}
\quad
\inferrule*
{
  \inc[\theta]{\Theta}{\Theta'} \\
}
{\inc[\theta]{\Theta,x:P}{\Theta',x:\theta P,\Xi}}
\quad
\inferrule*
{
  \inc[\theta]{\Theta}{\Theta'} \\
}
{\inc[\theta]{\Theta,\alpha}{\Theta',\alpha,\Xi}}
\end{mathparshrink}
A metasubstitution $\inc[\theta]{\Theta}{\Theta'}$ is a finite
map $\theta$ from flexible variables in $\Theta$ to well-formed
monotypes in $\Theta'$ such that the structure and dependency order of
$\Theta$ are preserved.
We write $\inc[\iota]{\Theta}{\Theta'}$ for an identity
metasubstitution and sometimes omit the name $\iota$.
For an identity metasubstitution, all entries of flexible variables in
$\Theta$ are preserved in $\Theta'$ but possibly with more monomorphic
information (e.g., from $\hat\alpha$ to $\hat\alpha\assign\tunit$).
Our algorithms always give identity metasubstitutions.
We also define equivalence of metasubstitutions
$\inc[\theta\equiv\theta']{\Theta}{\Theta'}$ standardly in
\Cref{app:metasubstitutions}.

All declarative judgements, including well-formedness, typing,
skeleton inference, skeleton refinement, subtyping, and consistency,
in \Cref{sec:declarative} are naturally extended to algorithmic
contexts.
We always retrieve the definitions of flexible variables from the
algorithmic context and explicitly write out type equivalence
judgements where we assumed them implicitly.
We show the changed rules in \Cref{app:decl-in-algo-context}.

\subsection{From Colours to Input Skeletons}
\label{sec:remove-colours}

Our declarative judgements for typing $\typp[\M]{\Gamma}{M}{\bid{Q}}$,
skeleton inference $\patp[\M]{\Gamma}{M}{\bid{Q}}$, subtyping
$\typsub[\N][\M][]{\Gamma}{\syn{P}}{\bid{Q}}$, and consistency
$\typconsis{\Gamma}{\syn{P}}{\bid{Q}}$ all use a bidirectional
skeleton $\bid{Q}$ on the RHS, whose colours distinguish between
inherited and synthesised information. (In typing, we know $\bid{Q}$
is a concrete type. We use $\bid{Q}$ here for uniformity with the
other three judgements.)

We do not have colours in our algorithmic judgements.
Instead, for each judgement, we add an input skeleton $P_0$ to
represent the inherited information.
As we will explain in detail in the following sections, our
algorithmic judgements for typing
$\atypp[\M][P_0]{\Theta}{M}{Q}{\Theta'}$, skeleton inference
$\apatp[\M][P_0]{\Theta}{M}{Q}{\Theta'}$, subtyping
$\atypsub[\N][\M][P_0]{\Theta}{P}{Q}{\Theta'}$, and consistency
$\aconsis[P_0]{\Theta}{P}{Q}{\Theta'}$ all have an input
skeleton $P_0$ and an output skeleton $Q$.
The synthesised information is the part of the output skeleton $Q$
that is not in $P_0$.
This is formally characterised by the refinement judgement
$\inh{P}\fillsyn\bid{Q}$ in \Cref{sec:skeleton}.
In the following sections, we use this judgement to establish the
connection between the declarative and algorithmic judgements in the
metatheory.

\subsection{Algorithmic Type Inference}
\label{sec:algo-type-inference}

\Cref{fig:selected-alg-type-inference} shows representative type inference
rules together with auxiliary skeleton splitting rules. The full type
inference rules are given in \Cref{app:algo-type-inference} and
correspond one-to-one with syntax-directed typing rules in
\Cref{sec:syntax-directed-typing}.
The type inference judgement has the form
$\atypp[\M][P]{\Theta}{M}{A}{\Theta'}$.
The input includes a context $\Theta$, a mode $\M$, a skeleton $P$,
and a term $M$; the output includes a type $A$ and a context
$\Theta'$.
As usual, we require all inputs to be well-formed.
In addition, we require a compatibility relation between modes and
input skeletons as a well-formedness condition, playing the same role
as the relation $\skmod[\TY]{\bid{A}}{\M}$ in \Cref{sec:typing}.
The relation $\skmod[\TY]{P}{\M}$ is defined as
${\skmod[\TY]{\Hole}{\TI}}$,
${\skmod[\TY]{A}{\TC}}$,
and
${\skmod[\TY]{P\to Q}{\FN\M}}$ if 
$\skmod[\TY]{Q}{\M}$.

\begin{figure}[tb] \small
\raggedright
\boxed{\atypp[\M][P]{\Theta}{M}{A}{\Theta'}}
\hfill
\begin{mathparshrink}
\inferrule*[Lab=\tilab{Abs}]
{
  \text{$\M\neq\TC$} \quad
  \mov{\Theta_0}{\split[P_0][P\to Q]{}}{\Theta_1} \quad
  \guessMono{P} = (\ol{\hat{\alpha}}; A) \\\\
  \atypp[\M-][Q]{\Theta_1, \ol{\hat{\alpha}}, x : A}{M}{B}{\Theta_2,x:A,\Xi}
}
{\atypp[\M][P_0]{\Theta_0}{\lambda x .M}{ A\to B}{\Theta_2,\Xi}}

\inferrule*[Lab=\tilab{AbsCheck}]
{
  \mov{\Theta_0,\ol\alpha}{\split[A_0][A\to B]{}}{\Theta_1,\ol\alpha} \\\\
  \atypp[\TC][B]{\Theta_1,\ol{\alpha},x : A}{M}{B}{\Theta_2,\ol{\alpha},x:A,\Xi} \\
}
{\atypp[\TC][\forall\ol{\alpha}.A_0]{\Theta_0}{
  \lambda x .M}{\forall\ol{\alpha}. A\to B}{\Theta_2}}

\inferrule*[Lab=\tilab{App}]
{
  \apatp[\TC][\Hole]{\Theta_0}{N}{P}{\Theta_1} \\
  \atypp[\FN\M][P\to Q]{\Theta_1}{M}{A \to B}{\Theta_2} \\
  \atypp[\TC][A]{\Theta_2}{N}{A'}{\Theta_3} \\
}
{\atypp[\M][Q]{\Theta_0}{M\; N}{B}{\Theta_3}}
\end{mathparshrink}

\raggedright
\boxed{\mov{\Theta}{\split[P_0][P\to Q]{} }{\Theta'}}
\hfill
\vspace{-.7\baselineskip}
\begin{mathparshrink}
  \inferrule*
  { }
  {\mov{\Theta_0}{\split[P\to Q][P\to Q]{} }{\Theta_0}}

  \inferrule*
  { }
  {\mov{\Theta_0}{\split[\Hole][\Hole\to \Hole]{}}{\Theta_0}}

  \inferrule*
  {
    \mov{\Theta_0,\hat\beta,\hat\gamma}{
      \unify{\hat\alpha}{\hat\beta\to \hat\gamma}}{\Theta_1}
  }
  {\mov{\Theta_0}{\split[\hat\alpha][\hat\beta \to \hat\gamma]{} }{\Theta_1}}
\end{mathparshrink}
\caption{Selected algorithmic type inference rules and skeleton splitting rules.}
\label{fig:selected-alg-type-inference}
\end{figure}

Rule \tilab{Abs} infers the type of a lambda abstraction when not in
checking mode, corresponding to the rule \sdtylab{Abs} in
\Cref{sec:syntax-directed-typing}.
It uses the skeleton splitting judgement
$\mov{\Theta}{\split[P_0][P\to Q]{}}{\Theta'}$ to split the input
skeleton into a function shape.
Here, $\Theta$ and $P_0$ are inputs while $P\to Q$ and $\Theta'$ are
outputs.
The last case is the most interesting one, where we split a flexible
variable $\hat\alpha$ into a function type with two fresh flexible
variables $\hat\beta$ and $\hat\gamma$ as its domain and codomain.
Rule \tilab{Abs} also uses the meta-function $\guessMono{P} =
(\ol{\hat\alpha};A)$ to get a type $A$ by filling in each ghost in $P$
with a fresh flexible variable in $\ol{\hat\alpha}$ and removing
universal ghosts in $P$.
Rule \tilab{AbsCheck} deals with the checking mode, corresponding to
the rule \sdtylab{AbsCheck}.
Rule \tilab{App} infers the type of an application, corresponding to
the rule \dtylab{App} in \Cref{sec:typing}.
We first infer the skeleton $P$ of the argument, then use it as the
argument skeleton to infer the type $A\to B$ of the function, and
finally check the argument against type $A$.

We now state the metatheory of type inference.
Following \citet{GundryMM10}, we first define the notion of problems
and solutions of type inference, and then state soundness,
completeness, and optimality.
Our proofs can be found in \Cref{app:proof-type-inference}.

\begin{definition}[Problems and solutions of type inference] A problem
of type inference is a tuple $(\Theta_0;\M;P;M)$ where $\Theta_0$ is
well-formed, $P$ and $M$ are well-formed in $\Theta_0$, and
$\skmod[\TY]{P}{\M}$.
A solution to it is a tuple $(\inc[\theta_1]{\Theta_0}{\Theta_1};A)$
such that $\typp[\M]{\Theta_1}{M}{\bid{A}}$ where $\Theta_1\vdash
\inh{\theta_1 P}\fillsyn\bid{A}$.
Such a solution is optimal if for any other solution
$(\inc[\theta]{\Theta_0}{\Theta};B)$,
there exists a metasubstitution $\inc[\zeta]{\Theta_1}{\Theta}$ such
that $\inc[\theta \equiv \zeta\circ\theta_1]{\Theta_0}{\Theta}$
($\theta$ factors through $\theta_1$ with cofactor $\zeta$) and
$\Theta \vdash \tyequiv{\zeta A}{B}$.
\end{definition}

\begin{restatable}[Soundness of type inference]{theorem}{soundTypeInference}
  \label{thm:sound-type-inference}
  Given a type inference problem $(\Theta_0;\M;P;M)$, if
  $\atypp[\M][P]{\Theta_0}{M}{A}{\Theta_1}$, then
  $(\inc{\Theta_0}{\Theta_1};A)$ is a solution.
\end{restatable}

\begin{restatable}[Completeness and optimality of type inference]{theorem}{completeTypeInference}
  \label{thm:complete-type-inference}
  Given a type inference problem $(\Theta_0;\M;P;M)$, if there exists
  a solution $(\inc[\theta]{\Theta_0}{\Theta};B)$, then
  $\atypp[\M][P]{\Theta_0}{M}{A}{\Theta_1}$ gives its optimal solution
  $(\inc{\Theta_0}{\Theta_1};A)$.
\end{restatable}

\subsection{Algorithmic Skeleton Inference}
\label{sec:algo-skeleton-inference}

We show three representative skeleton inference rules below.
The full skeleton inference rules are in
\Cref{app:algo-skeleton-inference} and correspond one-to-one with
declarative skeleton inference rules in \Cref{sec:skeleton-inference}.
Its judgement has the form $\apatp[\M][P_0]{\Theta}{M}{P}{\Theta'}$.
The input includes a context $\Theta$, a mode $\M$, a skeleton $P_0$,
and a term $M$; the output includes a skeleton $P$ and a context
$\Theta'$.
We require $\skmod[\SK]{P_0}{\M}$ as a well-formedness condition,
whose only difference from $\skmod[\TY]{P_0}{\M}$ is for the checking
mode: $\skmod[\SK]{\Hole}{\TC}$.
\begin{mathparshrink} \small
\inferrule*[Lab=\pilab{Abs}]
{
  \M\neq\TC \\
  \mov{\Theta_0}{\split[P_0][P\to Q_0]{}}{\Theta_1} \\\\
  \apatp[\M-][Q_0]{\Theta_1, x : P}{M}{Q}{\Theta_2,x:P,\Xi}
}
{\apatp[\M][P_0]{\Theta_0}{\lambda x .M}{ P\to Q}{\Theta_2,\Xi}}

\inferrule*[Lab=\pilab{AbsCheck}]
{
  \apatp[\TC][\Hole]{\Theta_0, x : \Hole}{M}{Q}{\Theta_1,x:\Hole,\Xi}
}
{\apatp[\TC][\Hole]{\Theta_0}{
  \lambda x .M}{\TopHole. \Hole\to Q}{\Theta_1,\Xi}}

\inferrule*[Lab=\pilab{App}]
{
  \apatp[\TC][\Hole]{\Theta_0}{N}{P_0}{\Theta_1} \\
  \apatp[\FN\M][P_0\to Q_0]{\Theta_1}{M}{P \to Q}{\Theta_2} \\
}
{\apatp[\M][Q_0]{\Theta_0}{M\; N}{Q}{\Theta_2}}
\end{mathparshrink}

Algorithmic skeleton inference rules closely mimic algorithmic type
inference rules in \Cref{sec:algo-type-inference}.
Rule \pilab{Abs}, unlike rule \tilab{Abs}, does not guess
a monomorphic type from $P$ but directly puts $x:P$ into the context.
Introducing flexible variables for ghosts here would be problematic
because these ghosts may be matched with inherited polytypes later, as
in our motivating example in \Cref{sec:introduction}.
Rule \pilab{AbsCheck} introduces a universal ghost for the function
skeleton, similar to \plab{AbsCheck} in \Cref{sec:skeleton-inference}.
Rule \pilab{App} does not do type checking for the argument $N$,
similar to \plab{App}.%

Problems and solutions of skeleton inference, as well as soundness,
completeness, and optimality theorems, are defined similarly to those
of type inference in \Cref{sec:algo-type-inference}.
The main difference is that to characterise the optimality of skeleton
inference, we need to formally define when a skeleton is more precise
than another skeleton.
We write $\Theta\vdash\precise{P}{Q}$, read as $P$ is more precise than
$Q$, if $P$ contains exactly all quantifiers in $Q$, but may have
more universal ghosts, and fewer monomorphic types.
This is exactly what $\Theta\vdash\inh{P}\fillsyn\inh{Q}$ means.
Since $\inh{Q}$ is inherited, the refinement relation cannot fill $Q$
with more quantifiers, but can only fill with monotypes and remove
universal ghosts.%
\footnote{It may appear intriguing that we say $P$ is more precise
than $Q$ but actually $Q$ can have more monotypes than $P$. We use the
word ``precision'' not in terms of the quantity of information, but in
terms of how a skeleton characterises terms with fewer unnecessary
restrictions.}
We therefore directly define $\Theta\vdash\precise{P}{Q}$ as
$\Theta\vdash\inh{P}\fillsyn\inh{Q}$.
On types, precision coincides with type equivalence.
Our optimality theorem guarantees that algorithmic skeleton inference
always infers the most precise skeleton.
Full theorems and proofs can be found in
\Cref{app:algo-skeleton-inference,app:proof-skeleton-inference}.

\subsection{Algorithmic Consistency}
\label{sec:algo-consistency}

We first introduce algorithmic consistency before algorithmic
subtyping.
\Cref{fig:selected-algo-consistency} shows representative algorithmic
consistency rules; the full rules are in \Cref{app:algo-consistency}.
Its judgement has the form $\aconsis[P_0]{\Theta_0}{P}{Q}{\Theta_1}$.
The input includes a context $\Theta_0$, a skeleton $P_0$, and a
skeleton $P$; the output includes a skeleton $Q$ and a context
$\Theta_1$.
This judgement refines $P$ to $Q$ guided by $P_0$, which itself
refines to $Q$.
Intuitively, $Q$ can be understood as the join of $P$ and $P_0$,
viewing the refinement relation as a partial order.
When both $P$ and $P_0$ are concrete types, this judgement degenerates
to the standard unification of $P$ and $P_0$ in type inference.

Refinement disallows uses of bound variables in ghosts.
Thus, to enforce that $Q$ is a valid refinement of $P$ and $P_0$, we
augment the judgement with two input variable sets $\Delta_1$ and
$\Delta_2$ to track bound variables in $P_0$ and $P$, respectively.
The full judgement has the form
$\aconsis[\Delta_1\mid\Delta_2][P_0]{\Theta_0}{P}{Q}{\Theta_1}$.
Rules \ulab{ForallInh}, \ulab{ForallSyn}, and \ulab{UnivGhost1} add
bound variables to $\Delta_1$ or $\Delta_2$.
The names of rules show their correspondence to declarative
consistency rules in \Cref{sec:skeleton-subtyping-consistency}.
Rules \ulab{Ghost} and \ulab{GhostR} restrict the uses of variables
from $\Delta_1$ or $\Delta_2$ in the output skeleton, guaranteeing the
output skeleton is a valid refinement of both input skeletons.

Compared to declarative consistency, there are three new rules 
dealing with the cases where either or both sides are flexible
variables: \ulab{Flex}, \ulab{FlexL}, and \ulab{FlexR}.
They use auxiliary judgements to look up definitions of flexible
variables and move other dependencies as necessary.
These auxiliary judgements are standard~\citep{GundryMM10} and are
defined in \Cref{app:algo-consistency}.
There are also two new rules to deal with ghosts in the input
skeletons: \ulab{GhostR} and \ulab{UnivGhostR}.
Declarative consistency does not have these two rules because it has
no explicit input skeletons.

The lemmas and proofs of soundness and completeness of consistency are
in \Cref{app:algo-consistency,app:proof-consistency}.

\begin{figure}[htbp] \rulesize
\raggedright
\boxed{\aconsis[\Delta_1\mid\Delta_2][P_0]{\Theta_0}{P}{Q}{\Theta_1}}
\hfill
\begin{mathparshrink}
\inferrule*[Lab=\ulab{Ghost}]
{
  \Delta_2 \cap \ftv{P_0} = \emptyset
}
{\aconsis[\Delta_1\mid\Delta_2][P_0]{\Theta_0}{\Hole}{P_0}{\Theta_0}}

\inferrule*[Lab=\ulab{ForallInh}]
{
  \aconsis[\Delta_1,\alpha\mid\Delta_2,\alpha][P_0]{
    \Theta_0,\alpha}{P}{Q}{\Theta_1,\alpha}
}
{\aconsis[\Delta_1\mid\Delta_2][\forall\alpha.P_0]{\Theta_0}{
  \forall\alpha.P}{\forall\alpha.Q}{\Theta_1}}

\inferrule*[Lab=\ulab{ForallSyn}]
{
  \aconsis[\Delta_1\mid\Delta_2,\alpha][\TopHole.P_0]{
    \Theta_0,\alpha}{P}{Q}{\Theta_1,\alpha}
}
{\aconsis[\Delta_1\mid\Delta_2][\TopHole.P_0]{\Theta_0}{
  \forall\alpha . P}{\forall\alpha. Q}{\Theta_1}}

\inferrule*[Lab=\ulab{UnivGhost1}]
{
  \aconsis[\Delta_1,\alpha\mid\Delta_2][P_0]{
    \Theta_0,\alpha}{\TopHole. P}{Q}{\Theta_1,\alpha}
}
{\aconsis[\Delta_1\mid\Delta_2][\forall\alpha.P_0]{\Theta_0}{
  \TopHole . P}{\forall\alpha. Q}{\Theta_1}}

\inferrule*[Lab=\ulab{Flex}]
{
  \mov{\Theta_0}{\unifyvar{\hat\alpha}{\hat\beta}}{\Theta_1}
}
{\aconsis[\Delta_1\mid\Delta_2][\hat\beta]{\Theta_0}{\hat\alpha}{\hat\alpha}{\Theta_1}}

\inferrule*[Lab=\ulab{FlexL}]
{
  \nonflex{P_0} \quad
  \mov{\Theta_0\mid\cdot}{\assignvar{\hat{\alpha}}{P_0}}{\Theta_1}
}
{\aconsis[\Delta_1\mid\Delta_2][P_0]{\Theta_0}{\hat{\alpha}}{\hat\alpha}{\Theta_1}}

\inferrule*[Lab=\ulab{FlexR}]
{
  \nonflex{P} \quad
  \mov{\Theta_0\mid\cdot}{\assignvar{\hat\alpha}{P}}{\Theta_1}
}
{\aconsis[\Delta_1\mid\Delta_2][\hat\alpha]{\Theta_0}{P}{\hat\alpha}{\Theta_1}}
\\

\inferrule*[Lab=\ulab{GhostR}]
{
  \Delta_1 \cap \ftv{P} = \emptyset %
}
{\aconsis[\Delta_1\mid\Delta_2][\Hole]{\Theta_0}{P}{P}{\Theta_0}}

\inferrule*[Lab=\ulab{UnivGhostR}]
{
  \Theta_0\vdash \guarded{P} \\
  \nonflex{P} \\
  \aconsis[\Delta_1\mid\Delta_2][P_0]{\Theta_0}{P}{Q}{\Theta_1} \\
}
{\aconsis[\Delta_1\mid\Delta_2][\TopHole.P_0]{\Theta_0}{P}{Q}{\Theta_1}}
\end{mathparshrink}

\caption{Selected algorithmic consistency rules.}
\label{fig:selected-algo-consistency}
\end{figure}

\subsection{Algorithmic Subtyping with Constraint Collection and Solving}
\label{sec:algo-subtyping}

The algorithmic subtyping judgement has the form
$\atypsub[\N][\M][P_0]{\Theta}{P}{Q}{\Theta'}$.
The input includes a context $\Theta$, a sort $\N$, a mode $\M$,
a skeleton $P_0$, and a skeleton $P$; the output includes a
skeleton $Q$ and a context $\Theta'$.
This judgement upcasts $P$ to $Q$ guided by the input skeleton $P_0$,
which itself refines to $Q$.
In addition to well-formedness of all inputs, we require
$\skmod[\N]{P_0}{\M}$ as a well-formedness condition.
Here we only show the most interesting rule \silab{ForallL} as
follows; the full rules are in \Cref{app:algo-subtyping}.
\begin{mathparshrink}\small
\inferrule*%
{
  \M \neq \TI \quad %
  \guarded{P_0} %
  \gor P_0 = \Hole \quad %
  \hl{\looksub[\N][\M]{\alpha}{P}{P_0}{P_1}[\Xi]} \quad
  \Theta_0,\Xi \vdash P_1 \quad %
  \atypsub[\N][\M][P_0]{\Theta_0,\Xi}{P[P_1/\alpha]}{Q}{\Theta_1}
}
{\atypsub[\N][\M][P_0]{\Theta_0}{\forall\alpha.P}{Q}{\Theta_1}}
\end{mathparshrink}

The side conditions avoid overlapping with other rules.
We use the novel look judgement
$\looksub[\N][\M]{\alpha}{P}{P_0}{P_1}[\Xi]$
defined in \Cref{fig:selected-look} to extract polymorphic
information for $\alpha$ before instantiation.
The input includes a sort $\N$, a mode $\M$, and two skeletons $P$ and
$P_0$. %
The goal is to find a proper substitution for $\alpha$ in $P$ such
that the substituted $P$ can be upcast to some $Q$ compatible with
$P_0$.
The type variable $\alpha$ may appear only in $P$. We require this as
a well-formedness condition.
We do not need the context $\Theta_0$ since our goal here is only
to extract polymorphic information, while flexible variables in
$\Theta_0$ only carry monomorphic information.
The output includes a skeleton $P_1$ that carries all polymorphic
information we extract for $\alpha$, and a context $\Xi$ of fresh
flexible variables used in $P_1$.
We check well-formedness of $P_1$ in the context $\Theta_0,\Xi$ before
instantiating $\alpha$ with it.%

The look judgement replaces the guessing of decorated skeletons in
\sublab{ForallL} in \Cref{sec:subtyping} and
\Cref{sec:skeleton-subtyping-consistency}.
A look judgement first collects all constraints $\mC$ and fresh
flexible variables $\Xi$ for $\alpha$ using the constraint-collection
judgement $\looksub[\M]{\alpha}{P}{P'}{\mC}[\Xi]$.
A constraint list $\mC$ is defined as
  $\mC ~::=~ \cdot \mid \subcons{\M}{P} \mid \eqcons{P},\mC$
with subtyping constraints $\subcons{\M}{P}$ and consistency
constraints $\eqcons{P}$.
It then solves these constraints using the constraint-solving
judgement $\solve{\mC}{\Hole}{Q}$, where $\Hole$ is the empty input
skeleton and $Q$ is the result.
Finally, when $\N=\SK$, we use the solved $Q$ as the final output;
when $\N=\TY$, since we want a type for instantiation, we further
transform $Q$ to a type $A$ using the meta-function $\guessMono{Q} =
(\Xi';A)$ as defined in \Cref{sec:algo-type-inference}.

Constraint collection is separated into two judgement forms: one for
subtyping $\looksub[\M]{\alpha}{P}{P'}{\mC}[\Xi]$ and the other for
consistency $\lookeq[\M]{\alpha}{P}{P'}{\mC}[\Xi]$.
Both judgements try to find the skeletons in $P'$ that correspond to
appearances of $\alpha$ in $P$, and collect these constraints.
Their rules follow the algorithmic subtyping and consistency rules,
respectively.
We discuss representative rules here.
Rules \lslab{Var} and \lelab{Var} deal with the base cases when the LHS is
the type variable we look for.
We add the constraint $\subcons{\M}{P}$ or $\eqcons{P}$ to the output
constraints.
Rules \lslab{Arrow} and \lelab{Arrow} deal with arrows.
We recursively look at the argument and result components, and append
their constraints and flexible variables together.
Rule \lslab{ForallL} removes a top-level universal quantifier on the
LHS.
We do not need to instantiate $\beta$, since our goal is to extract
polymorphic information for $\alpha$, which the instantiation of
$\beta$ contributes nothing to.
Rule \lelab{FlexR} deals with a flexible variable $\hat\beta$ on the
RHS.
Since $\alpha\in\ftv{P}$, we know that $\alpha$ corresponds to some
monotype which carries no polymorphic information.
Thus, we allocate a fresh flexible variable $\hat\gamma$ and add the
constraint $\eqcons{\hat\gamma}$.

Constraint solving $\solve{\mC}{P}{Q}$ sequentially solves each
constraint $C$, starting from the input skeleton $P$ and producing the
output skeleton $Q$.
It calls two auxiliary judgements: ${\solvesub{P}{\M}{Q}{P'}}$ for
solving a subtyping constraint and ${\solveeq{P}{Q}{P'}}$ for solving
a consistency constraint.
Their rules again follow the algorithmic subtyping and consistency
rules, respectively.
We discuss representative rules here.
Rules \solsublab{Var} and \soleqlab{Var} deal with the base cases when
the LHS is a rigid variable.
We directly keep this rigid type variable instead of forcing both
sides to coincide, since our goal is to extract polymorphic
information while a rigid variable is monomorphic.
Rules \solsublab{Arrow} and \soleqlab{Arrow} deal with arrow shapes.
We recursively solve the argument and result components, and build the
output arrow skeleton from the results.
Rule \solsublab{ForallL} deals with a universal quantifier on the LHS,
which we simply keep in the output.
Rule \soleqlab{GhostL} deals with the base where the LHS is a ghost
and we simply keep the RHS in the output.

\begin{figure}[tb] \small
\raggedright
\boxed{\looksub[\N][\M]{\alpha}{P}{P'}{Q}[\Xi]}
\text{\gray{look}}
\hfill
\begin{mathparshrink}
\inferrule*
{
  \looksub[\M]{\alpha}{P}{P'}{\mC}[\Xi] \\
  \solve{\mC}{\Hole}{Q} \\
  (\Xi';A) = \guessMono{Q} \\
}
{\looksub[\TY][\M]{\alpha}{P}{P'}{A}[\Xi,\Xi']}

\inferrule*
{
  \looksub[\M]{\alpha}{P}{P'}{\mC}[\Xi] \\
  \solve{\mC}{\Hole}{Q} \\
}
{\looksub[\SK][\M]{\alpha}{P}{P'}{Q}[\Xi]}
\end{mathparshrink}

\raggedright
\boxed{\looksub[\M]{\alpha}{P}{P'}{\mC}[\Xi]}
\boxed{\lookeq[\M]{\alpha}{P}{P'}{\mC}[\Xi]}
\text{\gray{constraint collection}}
\hfill
\[\ba{l@{\qquad}l@{\qquad}l}
\inferrule*[Lab=\lslab{Var}]
{
  ~
}
{\looksub[\M]{\alpha}{\alpha}{P}{\subcons{\M}{P}}[\cdot]}

&
\inferrule*[Lab=\lslab{Arrow}]
{
  \lookeq{\alpha}{P_1}{P'_1}{\mC_1}[\Xi_1] \quad
  \looksub[\M]{\alpha}{P_2}{P'_2}{\mC_2}[\Xi_2] \\
}
{\looksub[\FN\M]{\alpha}{P_1\to P_2}{P'_1\to P'_2}{\mC_1,\mC_2}[\Xi_1,\Xi_2]}

&
\inferrule*[Lab=\lslab{ForallL}]
{
  \looksub[\M]{\alpha}{P}{P'}{\mC}[\Xi] \\
}
{\looksub[\M]{\alpha}{\forall\beta.P}{P'}{\mC}[\Xi]}
\\[2ex]

\inferrule*[Lab=\lelab{Var}]
{
  ~
}
{\lookeq{\alpha}{\alpha}{P}{\eqcons{P}}[\cdot]}

&
\inferrule*[Lab=\lelab{Arrow}]
{
  \lookeq{\alpha}{P_1}{P'_1}{\mC_1}[\Xi_1] \\
  \lookeq{\alpha}{P_2}{P'_2}{\mC_2}[\Xi_2] \\
}
{\lookeq{\alpha}{P_1\to P_2}{P'_1\to P'_2}{\mC_1,\mC_2}[\Xi_1,\Xi_2]}

&
\inferrule*[Lab=\lelab{FlexR}]
{
  P \neq \alpha \\
  \alpha\in\ftv{P} \\
  \fresh{\hat\gamma}
}
{\lookeq{\alpha}{P}{\hat\beta}{\eqcons{\hat\gamma}}[\hat\gamma]}
\ea\]

\raggedright
\boxed{\solve{\mC}{P}{Q}}
\text{\gray{constraint solving}}
\hfill
\begin{mathparshrink}
\inferrule*
{~}
{\solve{\cdot}{P}{P}}

\inferrule*
{
  \solvesub{P}{\M}{Q}{P'}
}
{\solve{\subcons{\M}{Q}}{P}{P'}}

\inferrule*
{
  \solveeq{P}{Q}{P_1} \\
  \solve{\mC}{P_1}{P'}
}
{\solve{\eqcons{Q},\mC}{P}{P'}}
\end{mathparshrink}

\raggedright
\boxed{\solvesub{P}{\M}{Q}{P'}}
\boxed{\solveeq{P}{Q}{P'}}
\gray{\text{individual constraint solving}}
\hfill
\[\ba{l@{\qquad\qquad}l@{\qquad\qquad}l@{\qquad}l}
\inferrule*[Lab=\solsublab{Var}]
{~}
{\solvesub{\alpha}{\M}{Q}{\alpha}}

&
\inferrule*[Lab=\solsublab{Arrow}]
{
  \solveeq{P_1}{Q_1}{P'_1} \\
  \solvesub{P_2}{\M}{Q_2}{P'_2}
}
{\solvesub{P_1\to P_2}{\FN\M}{Q_1\to Q_2}{P'_1\to P'_2}}

&
\inferrule*[Lab=\solsublab{ForallL}]
{
  \solvesub{P}{\M}{Q}{P'}
}
{\solvesub{\forall\alpha.P}{\M}{Q}{\forall\alpha.P'}}

&
\\[2ex]

\inferrule*[Lab=\soleqlab{Var}]
{~}
{\solveeq{\alpha}{Q}{\alpha}}

&
\inferrule*[Lab=\soleqlab{Arrow}]
{
  \solveeq{P_1}{Q_1}{P'_1} \\
  \solveeq{P_2}{Q_2}{P'_2}
}
{\solveeq{P_1\to P_2}{Q_1\to Q_2}{P'_1\to P'_2}}

&
\inferrule*[Lab=\soleqlab{GhostL}]
{~}
{\solveeq{\Hole}{Q}{Q}}

&
\ea\]

\caption{Look rules, together with selected constraint collection and solving rules.}
\label{fig:selected-look}
\end{figure}

As an example, consider the algorithmic subtyping judgement
\[\small
\atypsub[\TY][\FN\FN\TC][(\forall b . \Hole\to b)\to \Int\to\Int]{\Theta}{
  \forall a . a \to a
}{Q}{\Theta'}
\]
for some $Q$, $\Theta$, and $\Theta'$. Its look judgement has the form:
\[\small
\looksub[\TY][\FN\FN\TC]{a}{
  a \to a
}{(\forall b . \Hole\to b)\to \Int\to\Int}{\forall b . \Int\to b}[\cdot]
\]
The constraint collection process collects two constraints
$\eqcons{\forall b.\Hole\to b}$ and $\subcons{\FN\TC}{\Int\to\Int}$ by
first applying \lslab{Arrow} and then \lelab{Var} and \lslab{Var} for
the argument and result components, respectively.
The constraint solving judgement is derived as follows, with two
individual constraint solving judgements highlighted and numbered.
\[\small
\inferrule*
{
  \refn[1][\solveeq{\Hole}{\forall b.\Hole\to b}{\forall b.\Hole\to b}]
  \\
  \inferrule*
  {
    \refn[2][\solvesub{\forall b.\Hole\to b}{\FN\TC}{\Int\to\Int}{
      \forall b .\Int\to b}]
  }
  {
    \solve{\subcons{\FN\TC}{\Int\to\Int}}{
      \forall b.\Hole\to b}{\forall b .\Int\to b}
  }
}
{
  \solve{\eqcons{\forall b.\Hole\to b},\subcons{\FN\TC}{\Int\to\Int}}{
    \Hole}{\forall b .\Int\to b}
}
\]

For \refn[1], we solve it by \soleqlab{GhostL}.
For \refn[2], we solve it by \solsublab{ForallL}, \solsublab{Arrow},
and then \soleqlab{GhostL} and \solsublab{Var} for the argument and
result, respectively.

We define problems and solutions of subtyping, and then state the
soundness, completeness, and optimality of subtyping.
The proofs can be found in \Cref{app:proof-subtyping}.

\begin{definition}[Problems and solutions of subtyping]
  A subtyping problem is a tuple $(\Theta_0;\N; \M;P_0;P)$ where
  $\Theta_0$ is well-formed, $P_0$ and $P$ are well-formed in
  $\Theta_0$, and $\skmod[\N]{P_0}{\M}$.
  A solution to it is a tuple $(\inc[\theta_1]{\Theta_0}{\Theta_1};P_1)$
  such that $\typsub[\N][\M][]{\Theta_1}{\syn{\theta_1 P}}{\bid{P_1}}$ where
  $\Theta_1\vdash \inh{\theta_1 P_0}\fillsyn\bid{P_1}$.
  The solution is optimal if for any other solution
  $(\inc[\theta]{\Theta_0}{\Theta};Q)$, there exists a 
  metasubstitution $\inc[\zeta]{\Theta_1}{\Theta}$ such that $\inc[\theta
  \equiv \zeta\circ\theta_1]{\Theta_0}{\Theta}$ and $\Theta \vdash
  \precise{\zeta P_1}{Q}$.
\end{definition}

\begin{restatable}[Soundness of subtyping]{theorem}{soundSubtyping}
  \label{thm:sound-subtyping}
  Given a subtyping problem $(\Theta_0;\N;\M;P_0;P)$, if
  $\atypsub[\N][\M][P_0]{\Theta_0}{P}{Q}{\Theta_1}$, then
  $(\inc{\Theta_0}{\Theta_1};Q)$ is a solution.
\end{restatable}

\begin{restatable}[Completeness and optimality of subtyping]{theorem}{completeSubtyping}
  \label{thm:complete-subtyping}
  Given a subtyping problem $(\Theta_0;\N;\M;P_0;P)$, if there exists
  a solution,
  $\atypsub[\N][\M][P_0]{\Theta_0\!}{P\!}{\!P_1}{\Theta_1}$ gives its optimal
  solution $(\inc{\Theta_0}{\Theta_1};P_1)$.
\end{restatable}

\FloatBarrier

\lstdefinestyle{metl}{
  language=[Objective]Caml,
  morekeywords={eff, handle, end, return, mask, forall},
  morecomment=[l]{--},
  literate=
    {->}{{$\to$}}2
    {=>}{{$\Rightarrow$}}2
    {()}{{()}}2
}

\section{Implementation and Extensions}
\label{sec:implementation}

We implement a prototype of \Calc with two main extensions: algebraic data
types and modal effect types (\Met)~\citep{TangWDHLL25,TangL26}.
The extension with algebraic data types is routine, but demonstrates that \Calc
smoothly scales to support standard type system features.
The extension with modal effect types is more interesting, as type inference for
first-class modal types faces similar challenges to type inference for FCP.
This extension demonstrates that \Calc scales to support further advanced type
system features.

The full source code of our prototype is available online~\citep{metl-implementation}, together with a range of examples and tests
including those from the \Met papers.
We leave the formal specification and proofs of our extensions to
future work.
We also provide a faithful encoding of the declarative type system
from \Cref{sec:declarative} in the Rocq theorem prover and verify that the
syntax-directed typing rules in \Cref{sec:syntax-directed-typing} are equivalent
to the declarative typing rules in \Cref{sec:typing}.

\subsection{Algebraic Data Types}
\label{sec:datatypes}

Our prototype implements algebraic data types and pattern matching.
This extension poses no particular challenge.
To illustrate how the declarative specification of \Calc can be
extended with data types as well, we show the skeleton inference rules
for pairs as follows.
\[\small\ba{l@{\quad}l@{\quad}l}
\inferrule*[Lab=\plab{Pair}]
{
  \patp[\TI]{\Gamma}{M}{\syn{P}} \quad
  \patp[\TI]{\Gamma}{N}{\syn{Q}} \\
}
{\patp[\TI]{\Gamma}{(M,N)}{\syn{P}\times \syn{Q}}}

&
\inferrule*[Lab=\plab{PairCheck}]
{
  \patp[\TC]{\Gamma}{M}{\syn{P}} \quad
  \patp[\TC]{\Gamma}{N}{\syn{Q}} \\
}
{\patp[\TC]{\Gamma}{(M,N)}{\TopHole.\syn{P}\times \syn{Q}}}

&
\inferrule*[Lab=\plab{Case}]
{
  \pati{\Gamma}{M}{\syn{\forall\ol{\alpha} [\TopHole]}. \syn{P_1} \times \syn{P_2}} \\\\
  \patp[\M]{\Gamma,x:\forall\ol{\alpha}[\TopHole].P_1,y:\forall\ol{\alpha}[\TopHole].P_2}{
    N}{\bid{Q}} \\
}
{\patp[\M]{\Gamma}{\Case M \Of \{(x,y) \mapsto N\}}{\bid{Q}}}
\ea\]
We have two rules for pair introduction, one for inference mode and
the other for checking mode.
In \plab{PairCheck}, we introduce a universal ghost to cover potential
type abstraction, similar to \plab{AbsCheck} in
\Cref{sec:skeleton-inference}.
We use a $\Casey$ construct to eliminate pairs.
We allow the term to be inferred with a polymorphic pair skeleton,
where we write $[\TopHole]$ for an optional universal ghost.
The branch term $N$ has the same mode $\M$ as the whole $\Casey$ term.

\subsection{Modal Effect Types}

Our prototype implements algebraic effects and
handlers~\citep{PlotkinP03,PlotkinP13}, and extends \Calc to do type
inference for modal effect types, a recent type system that
tracks the use of effects via modal types.
Our extension supports inference for both implicit introduction and
elimination of modalities, as well as implicit modal instantiation,
i.e., instantiation with modal types.

As observed by \citet{TangWDHLL25}, inferring implicit introduction
and elimination of modalities is closely related to inferring implicit
type abstraction and instantiation for FCP.
For instance, in \Met, a type $\forall a . \aeq{\Log}(a \to a)$
denotes a polymorphic function that may use the $\Log$ effect, as
specified by the modality $\aeq{\Log}$.
Consequently, for the type checking judgement
$\typp[\TC]{\Gamma}{\lambda x . x}{\forall a . \aeq{\Log}(a \to a)}$,
we infer not only the type abstraction of $a$, but also the modality
introduction of $\aeq{\Log}$.
As another example, for the application $\dec{run}\;\dec{idlog}$
where $\dec{run} : (\Int\to\Int)\to\Int$ and $\dec{idlog} :
\forall a . \aeq{\Log}(a\to a)$, we not only infer the instantiation
of $a$ to $\Int$, but also the elimination of the modality
$\aeq{\Log}$.
Usual bidirectional type systems, such as the one proposed in
\citet{TangWDHLL25}, can neatly infer such implicit introduction and
elimination of modalities, but only in a simply-typed setting.

Incorporating modal instantiation is challenging, for the same reason that
incorporating polymorphic instantiation in FCP is.
We must avoid guessing arbitrary modal types for instantiation.
Otherwise, for instance, $\dec{single}\;\dec{ilog}$ where
$\dec{ilog}:\aeq{\Log}(\Int\to\Int)$ would have two possible types
$\List\;(\aeq{\Log}(\Int\to\Int))$ and $\aeq{\Log}(\List\;(\Int\to
\Int))$ with no best choice.
We extend \Calc to infer modal types, making use of mixed directions
of information flow.
Mixed directions of information flow are particularly desirable for \Met.
A typical pattern of effectful programming is that given a polymorphic
effect handler $h : \forall a . \aex{E}(\tunit\to A)\to B$ for some
effect $E$ and types $A$ and $B$ (the modality $\aex{E}$ attaches to
the argument type), we want to apply this handler to a suspended
computation $\lambda x . M$ to handle the effects used in $M$.
In order to type $h\;(\lambda x . M)$, the information should first
flow from the argument $\lambda x. M$ to the function $h$ for
instantiation of $a$, then flow back to the argument to introduce the
modality $\aex{E}$.

Our extension of \Calc with \Met exploits much of the existing \Calc
infrastructure.
Universal ghosts $\TopHole$ now not only represent unknown
quantifiers, but also unknown modalities.
One key difference is that the \Met typing judgements track effect contexts that
are manipulated by modalities.
We add effect contexts as input to type inference judgements, but not
to skeleton inference judgements, since a universal ghost
represents unknown manipulation on effect contexts.
Another difference is that \Met has a kind system with two kinds.
We ignore kind checking in skeleton inference, since our goal is only
to collect polymorphic and modal information.
We do not refine kind inference for flexible variables in type inference, as
this is an orthogonal problem; instead, we simply assume the default kind when
insufficient kind information is provided by the context.
Another difference is that the accessibility of variables in a modal type system
is controlled by locks in the context.
We follow the right-residual approach of \citet{TangWDHLL25} for tracking
variable use.

\FloatBarrier
\section{Related and Future Work}
\label{sec:related-work}

Type inference for FCP spans a wide design space.
We organise the discussion around the two central questions we care
about when designing type inference for FCP: how to mix different
directions of information flow, and how to represent unknown type
information beyond unification variables.

\paragraph{Mixed Directions of Information Flow}
The line of work closest to \Calc does type inference for FCP by
considering information flow between functions and arguments.
We discuss the limitations of these approaches in terms of mixed
information flow and explain why they fail to type check our
motivating example in \Cref{sec:introduction}.
Boxy types~\citep{VytiniotisWJ06} merge the checking and inference
judgements of bidirectional typing into a single judgement in which
``boxes'' mark the parts of a type that must be inferred, and share
with \Calc the principle that polymorphism should never be guessed.
In the core system, information only flows from functions to
arguments.
They proposed several ad-hoc rules to allow mixing of information
flow, but none of these rules allow information to flow from one
argument to a sibling argument, so challenge (2) still fails.
GI~\citep{SerranoHVJ18} works in the opposite direction: it treats an
$n$-ary application as a whole and instantiates a variable
with polytypes only when the variable is \emph{guarded} under a type
constructor in some argument type.
GI does not allow information to flow from functions to arguments, so
challenge (2) fails.
Similarly, HMF~\citep{Leijen08} does not allow information to flow
from functions to arguments as robustly as \Calc does, so challenge
(2) fails.
QuickLook~\citep{SerranoHJV20} combines the two directions by first
having a quick look at the arguments of an application spine to guide
the polymorphic instantiation of the function. Then the type
information of the function is propagated back into the arguments.
As discussed in \Cref{sec:introduction}, although QuickLook addresses
challenge (2), it fails to address challenge (1) because it only
learns from an argument when the argument type is guarded, following
GI.
However, in our example, the argument $\churchIds$ has an unguarded
type $\ChurchList\;(\forall a.a\to a)$.
Moreover, the quick look process itself is a deliberately shallow
algorithmic pass aimed at easy implementation in GHC.
Skeleton inference of \Calc provides a more systematic and precise way
to collect information from arguments.

\paragraph{Local Type Inference}
The line of work on local type inference also cares about information flow.
As its name suggests,
local type inference (LTI)~\citep{PierceT00} combines bidirectional
checking with \emph{local} synthesis of type arguments.
However, LTI only allows information flow from arguments to functions,
and propagates information in an all-or-nothing fashion.
Coloured local type inference (CLTI)~\citep{OderskyZZ01} lifts the second
limitation by propagating partial types, with a colour on every type
component recording whether it is inherited or
synthesised.
Since everything is coloured, CLTI is not compatible with unification
of monotypes where the direction of information flow of monotypes is
left open until a solution is found.
\Calc works well with monotype unification by only ascribing colours
to universal quantifiers.
As discussed in \Cref{sec:overview-subtyping}, both CLTI and Boxy
types also restrict polymorphic instantiation more than \Calc does,
requiring every appearance of a guessed instantiation to be
determined.
Local Contextual Type Inference (LCTI)~\citep{XueCJO26} redesigns LTI
on top of contextual typing~\citep{XueO24}, using modes that reflect
program structure to guide instantiation. 
LCTI lacks colours as in \Calc and thus cannot represent fine-grained
information flow, where only part of the type information flows from
an argument to a function while the rest flows back.
Moreover, LCTI never instantiates the type of an argument and is
sensitive to argument orders.
On our example, $\churchIds$ can guide the instantiation of
$\churchMap$, but the polytype learned this way is not used when
checking the sibling lambda, which fails to solve challenge (2).

\paragraph{Beyond Unification Variables}
Every inference algorithm must handle type information that is not
yet known.
For monotypes, unification variables together with a simple solving
strategy suffice: equality constraints may be solved eagerly, in any order, without losing
generality.
This is not enough for general FCP, or any other scenario
where the solution does not form an equivalence class.
Committing to an instantiation too early can exclude valid
typings, so the order of solving starts to matter.
To address this, people have proposed different \emph{placeholders} to
represent what is known about an unknown type, and/or designed smart
\emph{solving orders} to delay work on an unknown type until it
becomes determined.
ATIA~\citep{Morris26} interleaves an Algorithm-M-style traversal that
pushes expected types towards the leaves with constraint solving, and
extends the syntax of types with \emph{pseudotypes} $[S]P$ that stand
for the result of an unknown sequence of instantiations.
\citet{PottierR06} stratify type inference for GADTs into a
constraint-based core preceded by a \emph{shape} inference phase,
where shapes are types with holes that propagate the information
contained in explicit annotations.
In a dependently typed setting with implicit arguments, 
FCIF~\citep{Kovacs20} elaborates into
a core theory with strictly curried telescope function types, so that
types themselves can defer decisions about inserting implicit
abstractions and applications.
Instead of enriching the placeholders, its successor
DOE~\citep{Kovacs26} postpones any checking problem whose expected
type is an undetermined metavariable and resumes it once the
metavariable is solved.
OmniML~\citep{obrien2025omniml} similarly suspends constraints
until the \emph{shape} of a type is uniquely determined by the
context, making the order of information flow dynamic rather than
fixed.
\citet{Parreaux26} proposes to go further and decouple such
resolution steps from type inference altogether.
None of the approaches described above leak placeholders into the
surface syntax, in contrast to approaches that enrich the type
language itself~\citep{BotlanR03,Leijen09,ParreauxBFC24}.
Most of these approaches also emphasise describing their algorithms
with their own placeholders and solving strategies, with no
declarative specification.
\citet{LeijenY25} embrace this trend and argue that specifications
should move \emph{closer} to algorithms, rephrasing Hindley-Milner
as inference under a prefix.
Similar to the above approaches, skeletons and ghosts of \Calc also
represent unknown type information for polymorphism.
In contrast to them, \Calc insists on a declarative specification of
the type system with soundness and completeness results.
\Calc gives a declarative specification of how skeletons for terms are
inferred.

\paragraph{Other Related Work}
Our ordinary ghosts $\Hole$ are reminiscent of the unknown type of
gradual typing~\citep{Siek2006GradualTF,CastagnaLPS19} and the
classical notion of type holes, while our universal ghosts $\TopHole$
are relatively new.
Our use of tags for polymorphic instantiation in \Cref{sec:subtyping}
to track provenance of guessed quantifiers is reminiscent of the work
on tracking the sources of type
errors~\citep{BhanukaPBB23,ZhaoMDBPO24,GuanYWH26}.

\paragraph{Future Work}
\label{sec:conclusion}

In addition to the ideas discussed in \Cref{sec:overview-limitations}
and formalising our extension of \Calc with modal effect types, other
directions for future work include: extending term-level type
annotations to skeleton annotations; extending \Calc to support deep
subtyping and let generalisation; and generalising \Calc to infer
other advanced type system features, such as GADTs, higher-kinded
types, and intersection and union types.

\FloatBarrier
\bibliographystyle{ACM-Reference-Format}
\bibliography{reference}

\includepdf[
  pages=-,
  pagecommand={\AppendixPDFPageCommand}
]{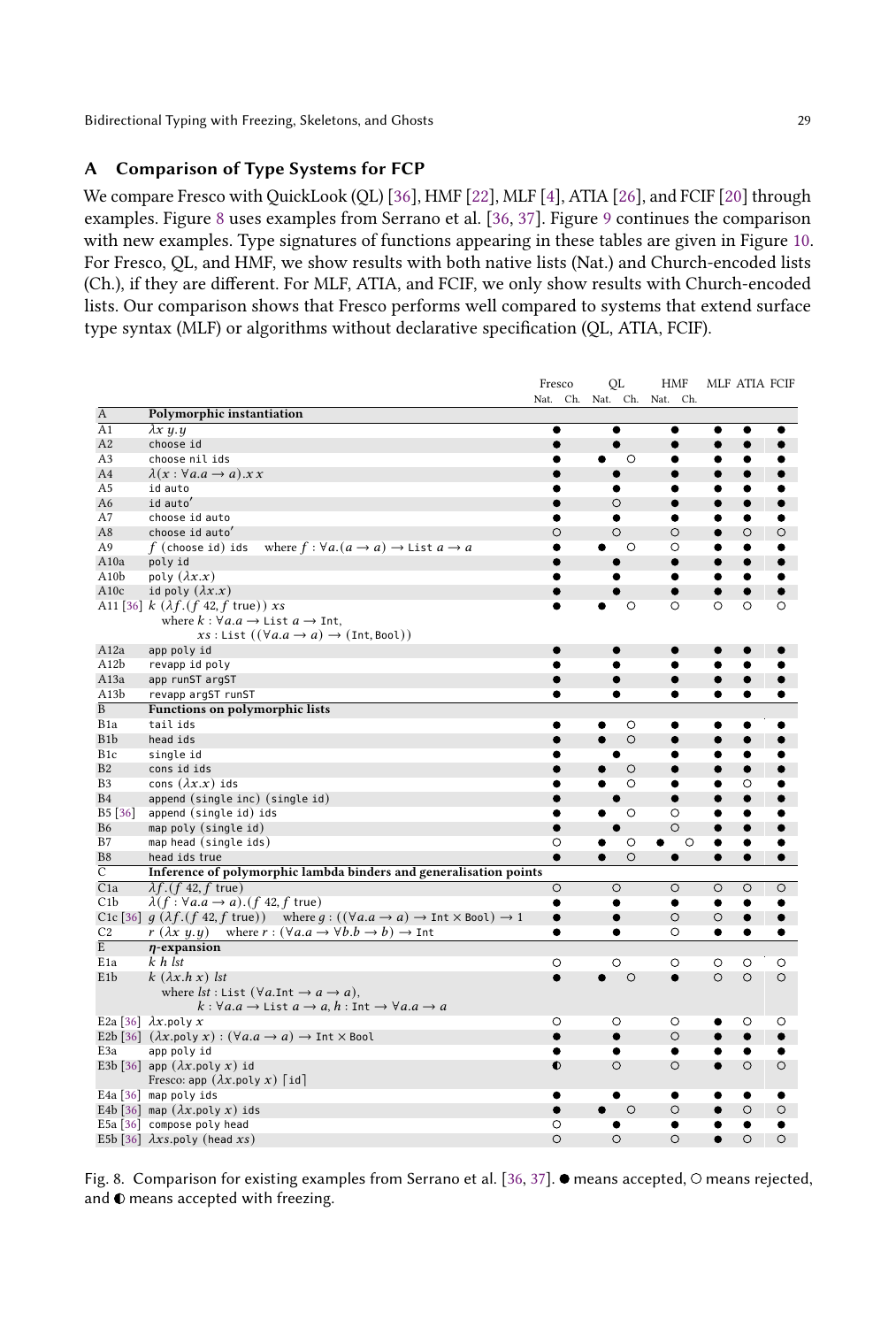}

\end{document}